\documentclass[12pt,letterpaper]{article}
\pdfoutput=1
\usepackage{jheppub}
\usepackage{xcolor}
\usepackage{graphicx}
\usepackage{wrapfig,enumerate,slashed}

\usepackage{footmisc}
\usepackage{amsmath}
\usepackage{wasysym} 
\usepackage{graphicx}
\usepackage{color}
\usepackage{comment}
\usepackage{hyperref}

\usepackage{amssymb}
\usepackage{amsmath}
\usepackage{epstopdf}
\usepackage{amsmath}
\usepackage{multirow}

\usepackage{url}
\usepackage{color}
\usepackage{listings}

\renewcommand{\d}{{\mathrm{d}}}

\DeclareRobustCommand{\Fig}[1]{Fig.~\ref{#1}}

\DeclareRobustCommand{\Eq}[1]{Eq.~(\ref{#1})}

\newcommand{\nn}{\nonumber} \renewcommand{\bf}{\textbf}
 
\renewcommand{\d}{\mathrm{d}}

\newcommand{\Kahler}{K\"ahler }

\definecolor{red1}{cmyk}{0,1,1,0.1}
\definecolor{blue1}{cmyk}{1,0,0,0}

\newcommand{\be}{\begin{equation}}
\newcommand{\ee}{\end{equation}}
\newcommand{\bea}{\begin{eqnarray}}
\newcommand{\eea}{\end{eqnarray}}

\newcommand{\Log}[1]{\log\left( #1 \right)}

\begin{document}

%\title{~~~~~~~~~~~~ ........................ }
\title{Sgoldstino-less inflation and low energy SUSY breaking}

\author[a]{Riccardo Argurio,}
\author[b,c]{Dries Coone,}
\author[d]{Lucien Heurtier}
\author[b]{and Alberto Mariotti}

\affiliation[a]{Physique  Th\'eorique  et  Math\'ematique  and  International  Solvay  Institutes, \\ Universit\'e  Libre
de Bruxelles, CP231, B-1050 Brussels, Belgium}
\affiliation[b]{Theoretische Natuurkunde and IIHE/ELEM, Vrije Universiteit Brussel, \\ and International Solvay Institutes, 
Pleinlaan 2, B-1050 Brussels, Belgium}
\affiliation[c]{Van Swinderen Institute for Particle Physics and Gravity, University of Groningen,\\
Nijenborgh 4, 9747 AG Groningen, The Netherlands}
\affiliation[d]{Service de Physique Th\'eorique, Universit\'e Libre de Bruxelles, CP225,\\
B-1050 Brussels, Belgium}

\emailAdd{rargurio@ulb.ac.be, a.a.coone@rug.nl, lucien.heurtier@ulb.ac.be, alberto.mariotti@vub.ac.be}

\abstract{
We assess the range of validity of sgoldstino-less inflation in a scenario of low energy supersymmetry breaking. 
We first analyze the consistency conditions that an effective theory of the inflaton and goldstino superfields should satisfy in order to be faithfully described by a sgoldstino-less model.  Enlarging the scope of previous studies, we investigate the case where the effective field theory cut-off, and hence also the sgoldstino mass, are inflaton-dependent.
We then introduce a UV complete model where one can realize successfully sgoldstino-less inflation and gauge mediation of supersymmetry breaking,
combining the $\alpha$-attractor mechanism and a weakly coupled model of spontaneous breaking of supersymmetry.
In this class of models we find that, given current limits on superpartner masses, the gravitino mass has a lower bound of the order of the MeV, i.e.~we cannot reach very low supersymmetry breaking scales.
On the plus side, we recognize that in this framework, one can derive the complete superpartner spectrum as well as compute inflation observables, the reheating temperature, and address 
the gravitino overabundance problem. 
We then show that further constraints come from collider results and inflation observables. Their non trivial interplay seems a staple feature of phenomenological studies of supersymmetric inflationary models.
}

\preprint{ULB-TH/17-08}

\maketitle

%%%%%%%%%%%%%%%%%%%%%%%%%%%%%%%%%%%%%%%%%%%%%%%%%%%%%
%  new section
%%%%%%%%%%%%%%%%%%%%%%%%%%%%%%%%%%%%%%%%%%%%%%%%%%%%%

\section{Introduction}
\label{sec:intro}

Recently there has been renewed interest in models of inflation in supergravity 
\cite{Ferrara:2014kva,DallAgata:2014qsj,Linde:2014ela,Carrasco:2015uma,Kahn:2015mla,Kallosh:2015nia,Dudas:2015eha,Ferrara:2015tyn,Carrasco:2015iij,Garcia-Etxebarria:2015lif,Ferrara:2016een,Kallosh:2016gqp,Linde:2016bcz,McDonough:2016der,Aalsma:2017ulu}. 
The inflationary paradigm for the very early universe is indeed supported by more and more experimental evidence, and the high energy scales that it naturally involves beg for its embedding in supergravity, as a first step towards a more complete quantum treatment in string theory models. 

A technical new result that has sparked much of the recent activity is the
formulation of supergravity theories with non-linearly realized supersymmetry, the so called nilpotent supergravity  \cite{Farakos:2013ih,Antoniadis:2014oya,Bergshoeff:2015tra}, which makes crucial use of the nilpotent (goldstino) superfield
 to implement supersymmetry breaking~\cite{Volkov:1973ix,Rocek:1978nb,Lindstrom:1979kq,Casalbuoni:1988xh,Komargodski:2009rz}.
This new formulation has been shown to be able to accomodate effortlessly models of single field inflation in supergravity with (at first sight) an arbitrarily low scale of supersymmetry breaking in the vacuum. In such models, indeed, an idealistic decoupling of the sgoldstino degrees of freedom is assumed (encoded by the nilpotency condition), which otherwise interact non-trivially with the inflaton. Technically speaking, the use of constrained superfields thus appears to drastically simplify inflationary models.\footnote{Models where the sgoldstino itself is the inflaton exist \cite{Achucarro:2012hg,Borghese:2012yu,Ferrara:2016vzg,Ketov:2014qha}, though it has been more popular to consider an inflaton superfield together with a stabilizer superfield \cite{Kawasaki:2000yn}, which can cure most of the problems of SUGRA inflation, and/or together with a supersymmetry breaking sector \cite{Buchmuller:2014pla,Buchmuller:2015oma,Dudas:2015lga}.}

We immediately see that there is a potential tension in this perspective. This is related to the energy scales that are involved in the problem, namely those of inflation, of supersymmetry breaking and of its mediation to the visible sector (the latter two are typically distinct, except in a handful of strongly coupled models). In other words, one might be worried that the cut-off scale of the sgoldstino-less effective field theory might fall below the scale of the physics that it is supposed to describe, i.e.~inflation.

In \cite{Dudas:2016eej} a class of UV complete models characterized by a fixed scale was considered. It was shown that requiring the UV completion not to spoil the inflationary dynamics (i.e.~for a nilpotent effective description to be viable) imposes stringent restrictions on the parameters of the model. Eventually, the supersymmetry breaking scale turns out to be  quite large, very close to the scale one would expect in gravity mediated models. 

In this paper, we investigate whether it is possible to construct more general UV complete models of inflation (which mimic sgoldstino-less inflation) 
with a low supersymmetry breaking scale after inflation. Our proposal is to consider the scale of the UV complete model as being inflaton-dependent.
The implementation is done by positing a connection between the inflationary sector and the hidden
sector responsible for generating the supersymmetry breaking in the MSSM through e.g.~gauge mediation (GMSB) \cite{Giudice:1998bp}, realizing a unifying framework for 
inflation and low energy supersymmetry breaking.%
\footnote{There are models where the inflaton resides directly into the hidden sector \cite{Ibe:2006rc,Nakai:2010km,Hamaguchi:2009hy,Kamada:2011ec,Fukushima:2012ra,Fukushima:2013vxa}, though there seem to be difficulties in getting towards low supersymmetry breaking scales.
}

We consider models of inflation where
the field breaking supersymmetry plays the role of the stabilizer,
while the inflaton field is not directly involved in the determination of the supersymmetry breaking vacuum in the hidden sector.
The latter is typically characterized by some messengers whose supersymmetric mass sets the scale of (gauge) mediation.
In our scenario the inflaton couples directly to these messenger fields. Thanks to this interaction, the
messengers get large inflaton dependent masses during inflation, while their masses at the end of inflation can be small, and hence typical gauge mediation can occur.

The interest of this comprehensive approach is that the inflaton sector, as well as the MSSM superparticle spectrum and couplings can be completely specified.
First, we find that the consistency of the theory implies that the scale of supersymmetry breaking can be in the regime suitable for gauge mediation to dominate, though in our class of weakly coupled models it is still not possible to access the lowest scales without dragging the superpartner masses to unacceptably small values.

Moreover, one can
then address issues associated to the reheating and the dynamics after inflation, for instance the
gravitino overabundance problem, and relate them to the expected superpartner spectrum. In such a framework the experimental constraints from cosmology,
like the number of e-folds and the values of $r$ and $n_s$, can be supplemented with experimental constraints from collider physics, for instance the gluino mass bound from LHC. With some simplifying assumptions we will actually show that this combined investigation can reduce significantly the allowed parameter space of an inflationary model, leading to predictions for both inflation observables and for the
superpartner spectrum.
In particular,
since the inflaton couples with the MSSM only through 
loops of heavy messenger fields, its decay width is small and hence the reheating temperature is generically quite low. This implies that the spectral tilt $n_s$ will be on the low end of the window compatible with experimental data.

In section \ref{EFTgeneral} we review basics of sgoldstino-less models for inflation and we introduce the generic consistency conditions that this class of theories should satisfy.
In section \ref{TheModel} we introduce our illustrative model, where the inflationary dynamics is governed by the $\alpha$-attractor mechanism \cite{Kallosh:2014rga,Kallosh:2015lwa,Linde:2015uga,Carrasco:2015rva,Carrasco:2015pla}.
We revisit the $\alpha$-attractor predictions and we discuss the regime of validity of the effective field theory (EFT) approximation in this concrete example. (Some more details are in Appendix \ref{regimes}.)
In section \ref{UVmodel} we introduce a possible UV completion of the illustrative model, based on a supersymmetry breaking sector with gauge mediation. We first analyze the consistency conditions of the UV theory and recover bounds compatible with those discussed in the EFT analysis.
In section \ref{Pheno} we perform a phenomenological analysis of the model. We compute the reheating temperature and implement bounds from the gravitino overabundance problem, Big Bang nucleosynthesis and LHC constraints. 
Finally, we conclude in section \ref{conclu}.

\section{Effective field theory for sgoldstino-less inflation}\label{EFTgeneral}
Sgoldstino-less models of inflation \cite{Ferrara:2014kva,DallAgata:2014qsj,Linde:2014ela,Carrasco:2015uma,Kahn:2015mla,Kallosh:2015nia,Dudas:2015eha,Ferrara:2015tyn,Carrasco:2015iij,Garcia-Etxebarria:2015lif,Ferrara:2016een,Kallosh:2016gqp,Linde:2016bcz,McDonough:2016der,Aalsma:2017ulu} are characterized by the presence of a nilpotent chiral superfield $S^2=0$ \cite{Volkov:1973ix,Rocek:1978nb,Lindstrom:1979kq,Casalbuoni:1988xh,Komargodski:2009rz}, i.e.~the goldstino superfield, responsible for spontaneous supersymmetry breaking
and another chiral superfield $\Phi$ whose imaginary component (in our conventions) is the inflaton field, which exhibits a shift symmetry in order to make its potential viable for inflation \cite{Kawasaki:2000yn}. 
The nilpotency condition corresponds to the fact that the scalar component of the
chiral superfield $S$, i.e.~the sgoldstino, has been integrated out. The theory with the nilpotent superfield (and potentially with other constrained superfields) is then interpreted as an effective field theory
valid up to the energy scale of the sgoldstino. 

In this section, our aim is to take a step back and write an effective field theory (EFT) which includes the sgoldstino, but where we can estimate its mass and follow its decoupling at low energies.
The sgoldstino gets its mass through non renormalizable operators in the \Kahler potential, which can be for instance generated perturbatively in a weakly coupled UV completion.
Such physics can be simply captured in a class of models of inflation in supergravity which is characterized by a few arbitrary functions. We take the following \Kahler and superpotential
\bea
&& 
K = \mathcal{K} (\Phi,\Phi^{\dagger}) + S S^{\dagger} -\frac{(SS^{\dagger})^2}{\Lambda_{\text{eff}}^2} 
\label{EFT1} \\
&&
W = f(\Phi) S+ M_p h(\Phi)
\eea
In these expressions the \Kahler potential for $\Phi$ typically respects a shift symmetry which makes it independent of the imaginary component, to be eventually identified with the inflaton field.
It can be typically non canonical with Planck scale corrections, reducing to a canonical form when 
the lowest component $\phi$ is small, i.e.~after inflation.
$f(\Phi)$ and $h(\Phi)$ are arbitrary functions of the inflaton field and can include higher dimensional operators typically suppressed by the Planck scale.
They determine the scalar potential for the inflaton and hence should be chosen properly in order to have a viable single-field inflationary dynamics and 
a (meta)-stable vacuum at the end of inflation.
Moreover, these functions have to be tuned in order to obtain a vanishing(ly small) cosmological constant in the vacuum at the end of inflation.

The shift symmetry of the \Kahler potential for $\Phi$ guarantees that the scalar potential for the imaginary part of $\phi$ does not
have an exponential growth during inflation. On the other hand in order to generate a potential to drive inflation, the shift symmetry should be broken. The proposal in \cite{Kawasaki:2000yn} is to introduce a massive term in $f(\Phi)$ to break the shift symmetry softly. 
It is also possible to break the shift symmetry directly in the \Kahler potential, as long as it is still an approximate symmetry.

The validity of the sgoldstino-less description is controlled by the sgoldstino mass, which is $\phi$ dependent
\be \label{msg}
m_{s}^2 =\frac{4| f(\phi)|^2}{\Lambda_{\text{eff}}^2}\ .
\ee
The fermionic component of $S$ is the goldstino and it is massless. It is eaten by the gravitino which  then acquires the following mass 
\be
\label{m32}
m_{3/2} \simeq \frac{|W|}{ M_p^2}\ .
\ee
Note that the sgoldstino mass depends on another scale setting the validity of this effective theory which is $\Lambda_{\text{eff}}$.  By definition of EFT, this scale must be larger than $m_s$. One should then wonder what is the typical size of $\Lambda_{\text{eff}}$ with respect to the Planck scale and how 
such a non-renormalizable operator is generated in a UV completion of the theory. 
If $\Lambda_{\text{eff}} \sim M_p$  the simplest interpretation is that this quartic operator arises from Planck scale physics.
If instead $\Lambda_{\text{eff}} \ll M_p$ then one is necessarily integrating over some
physics below the Planck scale and the operator leading to the sgoldstino mass can be typically interpreted as the leading term in a series of higher dimensional corrections to the \Kahler potential
suppressed by powers of $\Lambda_{\text{eff}}$.

In the first case, the sgoldstino mass at the end of inflation will scale as $\frac{f_0}{M_p}$, where $f_0 \equiv f(0)$. In order for this scalar to be decoupled from SM physics we should demand its mass to be 
larger than roughly a TeV. This automatically sets $f_0$, the scale of supersymmetry breaking at the end of inflation, to be large, and poses the model in a scenario where the gravity mediated 
contribution to the MSSM soft terms, also scaling as $\frac{f_0}{M_p}$, are sizable. Actually, if $\Lambda_{\text{eff}} \sim M_p$, in the vacuum the gravitino mass turns out to be  of the same size as $m_s$. 

In the second case, the sgoldstino mass is given by $\frac{f_0}{\Lambda_{\text{eff}}} \gg \frac{f_0}{M_p}$. Then $f_0$ can be small keeping $m_s$ sizeable, and SUSY breaking can be mediated to the MSSM via gauge interactions, with 
gravity mediated effects subleading.
In this paper we are interested in the second case, and hence we are immediately facing questions associated to the validity of the effective theory at very high scales and the role played by $\Lambda_{\text{eff}}$.
In \cite{Dudas:2016eej} the authors discussed issues associated to UV completions of sgoldstino-less models and found that a viable scenario needs an effective scale $\Lambda_{\text{eff}}$ of at most one order of magnitude 
smaller than the Planck scale. Here we discuss systematically the consistency conditions that we expect the effective theory to satisfy, and we propose a strategy to realize models of inflation 
with low
supersymmetry breaking scale at the end of inflation. 

\subsection{Consistency conditions}
\label{condi}

Let us start with identifying the consistency conditions that the effective theory in (\ref{EFT1}) should satisfy. 
The idea is, again, that the EFT should be in its regime of validity at the energy scales of the physics that one is describing, that is during and after inflation.  
Hence, these conditions must be satisfied along the entire inflaton trajectory, 
from Planckian values to the origin:
\begin{equation}\label{conds}
m_{s}^2 \ll \Lambda_{\text{eff}}^2\ , \qquad 
\langle s \rangle \ll \Lambda_{\text{eff}}\ . 
\end{equation}
These two conditions are necessary to have a well defined effective theory for $S$. 
Observe that all quantities appearing in these conditions should be intended as functions of the inflaton VEV.
The first condition in \eqref{conds} is easily translated into the constraint
\be
\label{easy}
f(\phi) \ll \Lambda_{\text{eff}}^2\ , 
\ee
The second condition in \eqref{conds} is involved and requires the study of the sgoldstino VEV along the entire inflaton trajectory.

Note that strictly speaking, an EFT such as \eqref{EFT1} treats differently $\Phi$ and $S$, in the sense that integrating out the physics at $\Lambda_{\text{eff}}$ affects the \Kahler potential of $S$, while the one of $\Phi$ is more generally determined by Planck scale physics. Nevertheless, we can conservatively ask that also the degrees of freedom of $\Phi$, at least after inflation, are within the regime of validity of the EFT of $S$:
\begin{equation}\label{condphi}
m_{\phi}^2 \ll \Lambda_{\text{eff}}^2\ .
\end{equation}

Besides these consistency conditions, in order to guarantee that inflation is driven by a single field (or equivalently, that the goldstino superfield is nilpotent at that time), we should demand that during inflation the sgoldstino is heavier than the typical scale, i.e. the Hubble scale, hence
\begin{equation}\label{condnilp}
m_{s}^2 {\Large |}_{\text{infl}} \gg H^2 {\Large |}_{\text{infl}}\ .
\end{equation}
On the other hand
this translates simply to 
\begin{equation}
\Lambda_{\text{eff}} \ll M_p\ ,
\end{equation}
where we assume a potential dominated by $|f(\phi)|^2$. We thus see that having a scale for new physics lower than the Planck scale is actually a requirement for contemplating the decoupling of the sgoldstino from inflation.

Note that generically the effective scale $\Lambda_{\text{eff}}$ can  be a function of the inflaton field, varying along the inflationary trajectory. 
This is a promising possibility since a function $\Lambda_{\text{eff}}(\phi)$ increasing with $\phi$ implies that the validity threshold of the effective theory grows with increasing $\phi$.
Note that a $\phi$ dependent $\Lambda_{\text{eff}}$ would introduce a new source of shift symmetry breaking, which should be controlled by a small parameter.

In the next section we will  study the implications of the consistency conditions
on the effective field theory with dynamical $\Lambda_{\text{eff}}(\Phi)$ in a simple model.
We will then provide a UV completion which generates such effective theory.

\section{An illustrative model}\label{TheModel}

\subsection{Definition of the model}
The lagrangian we will consider in this section is of the form
\bea
&& \label{alphakahler}
K= - 3\frac{\alpha}{2} M_p^2 \log \left( \frac{(M_p^2 - \Phi \Phi^{\dagger})^2}{(M_p^2 +\Phi^2)(M_p^2 +\Phi^{\dagger 2})}  \right) + S S^{\dagger} -\frac{(S S^{\dagger})^2}{\Lambda_{\text{eff}}(\Phi)^2}\;, \\
&&
W = f(\Phi) S +M_p h(\Phi)\;, \nonumber
\eea
where the scale $\Lambda_{\text{eff}}$ is, as announced above, promoted to be a function of the inflaton field value during inflation. The choice of a non minimal form for the K\"ahler potential (following the so-called $\alpha$-attractor set up \cite{Kallosh:2014rga,Kallosh:2015lwa,Linde:2015uga,Carrasco:2015rva,Carrasco:2015pla}) for the inflaton $\Phi$ will allow us to reach much better observables while keeping polynomial and independent the functions $f$ and $h$, which will show some importance later on.

The three different functions introduced above will be taken of the form
\be
f(\Phi)=f_0 - \frac{m_f}{M_p} \Phi^2\,, \quad h(\Phi)=h_0 - \frac{m_h}{M_p} \Phi^2\,, \label{eq:fhFunctions}
\ee 
and
\be
\Lambda_{\text{eff}}^2(\Phi)=|\Lambda_0 + g \Phi|^2\,. \label{eq:LambdaFunction} 
\ee
In this situation, $\Lambda_{\text{eff}}^2(\Phi)$ will take large values during inflation (where $\text{Im}(\Phi) \sim M_p$) and fall down to lower values when $\Phi$ rolls down its potential.

Several remarks should be made about the structure of the \Kahler potential in \eqref{alphakahler}.
For field values much smaller than the Planck scale, the 
$\alpha$-attractor \Kahler potential is a canonical one (for $\alpha=1/3$) exhibiting a shift symmetry for $\text{Im}(\Phi)$. 
The non-renormalizable term for $S$ suppressed by $\Lambda_{\text{eff}}(\Phi)$ is instead a
non canonical term, whose $\Phi$ dependence introduces an extra source of breaking of the shift symmetry. 
Henceforth we consider $g \ll 1$ such that the breaking of the shift symmetry is small.
Moreover, the scale $\Lambda_{\text{eff}}(\Phi)$ represents the mass scale where we expect new physical states.
Hence we observe that we cannot explore regions where
$\text{Re}(\Phi) = -\frac{\Lambda_0}{g}$ since in this locus the \Kahler potential becomes singular, corresponding to some states in the UV completion of the model that become 
massless.

As already stated, in taking into account corrections to the \Kahler potential in Eq.~\eqref{alphakahler}, somehow we are considering on different footing the inflaton superfield $\Phi$ and the sgoldstino superfield. The assumption is that the $\alpha$-attractor type potential is set by Planck scale physics, while the corrections to the sgoldstino are generated at a much smaller scale. This assumption is valid as soon as 
$\Lambda_{\text{eff}}(\Phi) \ll M_{p}$, which we ensure by demanding $\Lambda_0 \ll M_{p}$ and $g \ll 1$. 
Note that $g\ll 1$ also ensures that condition \eqref{condnilp} is satisfied.

The states at the scale $\Lambda_{\text{eff}}(\Phi)$ will also generically generate \Kahler corrections
for the field $\Phi$, possibly mixing $\Phi$ and $S$. However, we keep only the corrections to the field $S$ since the sgoldstino would be massless at tree level if not for such corrections, which are the leading terms generating a sgoldstino mass. In the case of the inflaton field $\Phi$, possible \Kahler corrections to its mass from physics at the scale $\Lambda_{\text{eff}}(\Phi)$ will be proportional to $g$ and subleading with respect to the
tree level terms.
We will see how to generate the corrections to the \Kahler potential in a UV complete theory in the next section, where also other terms (allowed by the symmetries) will be present, but  will not change qualitatively the analysis of this section.

\subsection{EFT validity analysis}
\label{EFTanal}
We now proceed with the analysis of the scalar potential and the consistency conditions of the EFT description.
Going to a real basis, one can define
\be
\phi\;\equiv \; \frac{\chi+i \varphi}{\sqrt{2}}\,,\quad S\; \equiv\; \frac{s+i \sigma}{\sqrt{2}}\,,
\ee
where due to the structure of the K\"ahler potential, $\varphi$ will play the role of the inflaton in what follows. The absence of linear terms in a small field expansion in $\chi$ guarantees that the so called {\em sinflaton} is stabilized to zero VEV during the whole inflationary trajectory.

Assuming the validity of the effective formulation and that the stabilizer $S$ acquires a sufficient mass during inflation, which will be checked a posteriori, one can derive the inflaton-dependent vev of the scalar $s$ to be
\be 
\langle s \rangle\;=\;M_p\frac{(\varphi ^2-2M_p^2)^2 f' h'+24 \alpha  f hM_p^2}{6 \sqrt{2} \alpha  \left(-\frac{(\varphi ^2-2M_p^2)^2 }{12 \alpha }(f'^2+h'^2)+\frac{4 f^2M_p^4}{\Lambda_{\text{eff}}^2}-2 h^2M_p^2\right)}\,, \label{vevS}
\ee
where a prime denotes a derivative with respect to the complex field $\Phi$ in \eqref{eq:fhFunctions}, and the functions $f$, $h$ and $\Lambda_{\text{eff}}$ are all dependent on $\Phi = i\frac{\varphi}{\sqrt{2}}$ during inflation.

In the vacuum at the end of inflation, where $\Phi=0$, the cosmological constant can be removed by imposing that
\be V(\varphi=0)=0\,,\ee
which is equivalent to fixing
\be 
\label{cctuning}
h_0^2\;=\;\frac{f_0^2}{3}\left(1+\mathcal O (\frac{\Lambda_0^2}{M_p^2})\right)\,.
\ee
Note that in this vacuum $\langle s \rangle \propto \Lambda_0^2/M_p$, i.e.~it is of the same order as the corrections to the above relation.
In addition, the physical mass of the inflaton $m_{\varphi}$ is given by the parameter $2 m_h$, whereas the gravitino mass \eqref{m32} will fall down at the end of inflation to $f_0/\sqrt 3 M_p$.

The vacuum at vanishing $\Phi$ is not a global minimum. Indeed, 
there is a supersymmetric vacuum in field space where the sinflaton gets a VEV of order $\sqrt{\frac{f_0 M_p}{m_f}}$. One therefore has to ensure that there are no tachyonic directions about the 
extremum.
We find that
the stability of the non supersymmetric vacuum in the sinflaton direction imposes the constraint\footnote{Using the approximate formula $S_\mathrm{bounce}\propto (\Delta\phi)^4/\Delta V$, 
a qualitative estimate of 
the tunneling rate into the SUSY vacuum 
indicates that 
the metastable vacuum is sufficiently long-lived for the typical range of parameters that we will consider.}
\be
\label{notachion}
m_h^2 > \frac{3 \alpha f_0 m_f}{2 M_p} \ .
\ee 

The inflationary trajectory spans the space of $\text{Im}(\Phi)$ with $\text{Re}(\Phi) = \text{Im}(S) = 0$ while the real part of the sgoldstino is given by equation (\ref{vevS}).
We should follow the inflationary trajectory in the $\varphi-s$ space in order to verify that the conditions listed in the previous
sections (and the absence of tachyons) are satisfied.\footnote{We will not address the issue of the initial conditions here, rather we will assume the inflaton starts rolling at Planckian VEVs.}

The simple conditions \eqref{easy} and \eqref{condphi} are translated in the following constraints
\be
\label{easyconstraints}
f_0 < \Lambda_0^2\ , \qquad \frac{m_f}{M_p} < g^2 \ ,\qquad m_h < \Lambda_0\ .
\ee
We note that the coupling $g$ is crucial to keep the sgoldstino mass within the validity of the effective theory. We will see how these bounds correspond typically to no-tachyon conditions in explicit UV completions of the model.

The third condition advocated in Section \ref{EFTgeneral} is
\be
\label{sgoldbis}
\langle s \rangle \ll \Lambda_{\text{eff}}(\Phi)
\ee 
and it is quite involved to solve analytically given the different terms entering into the expression for the sgoldstino VEV 
(\ref{vevS}).
In Appendix \ref{regimes} we perform a simplified analysis of this VEV and we find that imposing the following inequalities
\be
\label{regime_ine}
f_0  \ll \frac{\Lambda_0^2 m_f}{ M_p}\ ,   \qquad   \frac{\Lambda_0^2}{M_p^2} \ll \frac{m_h}{m_f}\ , \qquad \frac{\Lambda_0}{M_p} \gg \frac{m_h}{m_f} \ ,
\ee
one is guaranteed that the sgoldstino VEV is within the validity of the EFT along the whole inflationary trajectory.\footnote{Other valid regions in parameter space exist, but we found that this is the only one that survives the more stringent bounds of the weakly coupled UV completion that we will consider in the next section.}
 
To conclude, we find that in the model \eqref{alphakahler} the EFT is consistent all along inflation if the conditions \eqref{notachion}, \eqref{easyconstraints} and \eqref{regime_ine} are satisfied.
Note that the last condition in \eqref{regime_ine} leads automatically to $m_h \ll m_f$. This implies that during inflation, when the 
inflaton takes field values of order $M_p$, the parameter $m_f$ 
dominates the scalar potential and hence sets the scale of inflation. The mass of the inflaton at the end of inflation is instead controlled by $m_h$, which is a different 
(and much smaller) scale.

\subsection{$\alpha$-attractor inflation}

During inflation, the potential will be dominated by the mass scale $m_f$ which will be fixed by the 
cosmological observables. Indeed, in an expansion for small parameters $f_0$, $m_h$ and $\Lambda_0$, the effective potential during inflation is, at leading order
\be 
\label{eq_pot_simplyfied}
V_{\text{inf}} \;\approx\; \frac{m_f^2}{4M_p^2} \varphi^4\,.
\ee
However, due to the non-canonical \Kahler potential, we have to normalize the inflaton using
\be 
\varphi \;\equiv\; \sqrt{2}\ \tanh \left[\frac{\tilde \varphi}{\sqrt{6\alpha}M_p}\right]\,.
\ee
As usual in $\alpha$-attractor scenarios~\cite{Kallosh:2013yoa,Carrasco:2015rva,Kallosh:2014rga,Carrasco:2015pla,Kallosh:2015lwa,Linde:2015uga},
one can use the COBE normalization and the slow roll conditions to extract the values of $n_s$, $r$ and $m_f$ as a function of $\alpha$ and the number
of e-folds, denoted $N$.
We here focus on the case of the quartic power potential as in \eqref{eq_pot_simplyfied}.
Moreover, in order to provide analytic expressions, we can consider $N$ to be large and expand the relevant quantities in powers of
$1/N$.

The inflation observables $n_s$ and $r$ at second order in this expansion and in slow roll are \cite{Gong:2001he}
\begin{align}
n_s&=1-\frac{2}{N}+\frac{1}{N^2}\left(\frac{4}{3}-2C+\frac{\alpha}{2}\left(\frac{1}{2}\rho-3\right)\right), \\
r&=\frac{12\alpha}{N^2}\left(1+\frac{1}{N}\left(2C-\frac{1}{12}(8+3\alpha\rho)\right)\right), \qquad
\rho=\sqrt{9+\frac{48}{\alpha}}\ ,\label{eq_nsrN}
\end{align}
where $C=\gamma_E+\log(2)-2$ and $\gamma_E$ is the Euler constant.
Notice that for $\alpha = 1$ these relations, at first order in $1/N$, reproduce the Starobinsky and the Higgs inflation model, though the second order term is different \cite{Roest:2013fha}.

Moreover,
the COBE normalisation $A_s=(24\pi^2M_p^4)^{-1}V/\epsilon_*\approx2.2\cdot10^{-9}$ can be used to fix 
\begin{equation}
\label{mffix}
	m_f^2=A_s \frac{18\alpha\pi^2}{N^2}M_p^2,
\end{equation}
where we expanded $m_f$ only to first order in $1/N$ since we do not require additional accuracy.

\begin{figure}[th]
\begin{center}
\includegraphics[width=0.7\linewidth]{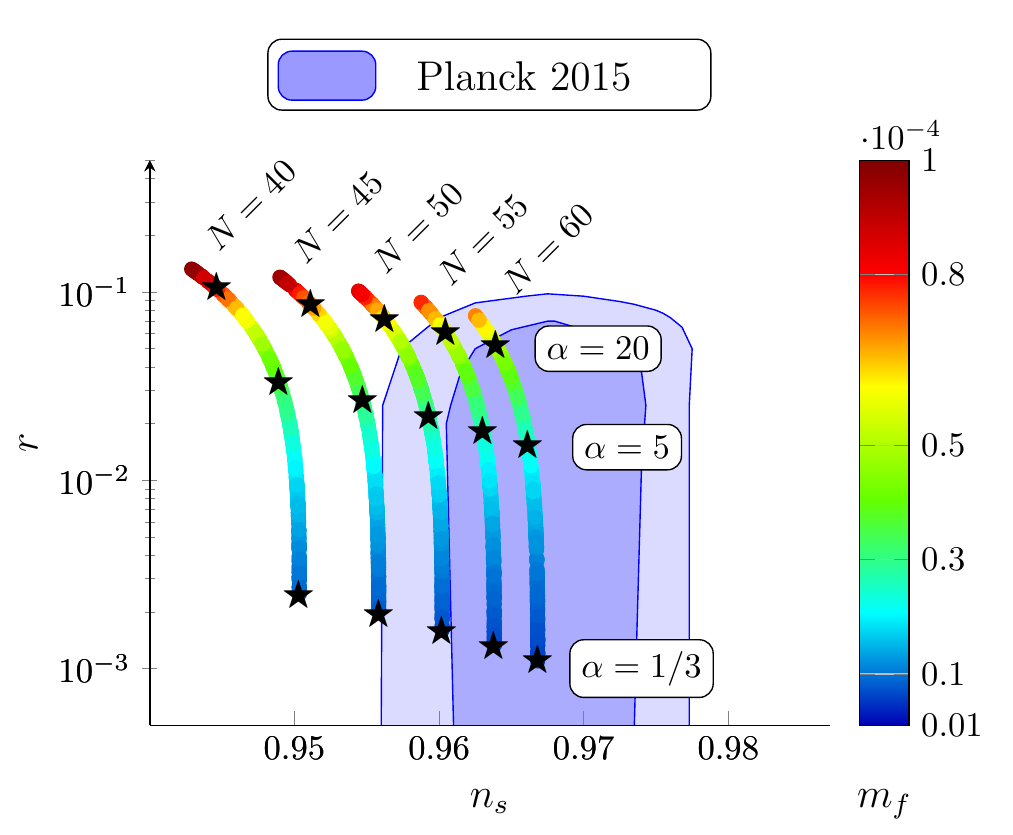}
\caption{\label{fig:alphaObs}Inflation observables for $\alpha$-attractor models using different values of the $\alpha$ parameter and number of e-folds. The color bar indicates the value of $m_f$ (in units of $M_p$) required for satisfying the COBE normalization measurement. The shaded region indicates the values favoured by the Planck collaboration at 1 and 2 $\sigma$ \cite{Ade:2015lrj}.}
\end{center}
\end{figure}

The predictions for the $\alpha$-attractor model with quartic potential is displayed in 
\Fig{fig:alphaObs} for different choices of $\alpha$ and of the number of e-folds $N$.
In  Section \ref{Pheno}, once we will have specified the UV completion, we will compute the reheating temperature and derive the
expected number of e-folds.
One can observe that anyway the required values of $m_f$ stand between $10^ {-6}M_p$ and $10^{-4}M_p$, the lowest values being favoured for $\alpha=1/3$.

Note that generically for small numbers of e-folds one can obtain from the $\alpha$-attractor models values of $n_s$ which are in the low-end region of the 
allowed Planck contours.

\subsection{Remarks on chaotic inflation}
Before proceeding to analyse a UV completion of this model, 
we would like to make a few remarks about the consequences of the
consistency conditions identified in this section for other inflation models.
In particular we would like to assess if nilpotent chaotic inflation (i.e.~with a quadratic inflaton potential) can be consistently realized since, even if it is disfavored by recent  measurements of the tensor-to-scalar ratio $r$ \cite{Ade:2015lrj,Ade:2015tva,Array:2015xqh}, it could represent
a minimal application of the nilpotent paradigm.

The most economical possibility would be to consider a shift symmetric \Kahler potential $\mathcal{K}=\frac{1}{2}(\Phi+ \Phi^{\dagger})^2$ and a quadratic expression for $h(\Phi)$,
with $h''(\Phi)$ setting the scale of inflation.
In order for the stabilization of the scalar potential to be effective at large field values, one should choose $f(\Phi)$ properly.
For instance, further requiring the cosmological constant in the vacuum to be zero restricts to the choice $f(\Phi)=\sqrt{3} h(\Phi)$ \cite{DallAgata:2014qsj,Kallosh:2014hxa,Scalisi:2015qga}.

This immediately leads to two important issues. 
First consider the case in which $\Lambda_{\text{eff}}$ is a constant independent on the inflaton VEV. Then  
the requirement $f(\phi) \ll \Lambda_{\text{eff}}^2$ already puts strong constraints on the allowed values for $\Lambda_{\text{eff}}$.
For instance, if $f(\phi)$ is a quadratic function, i.e. $f(x)=f_0-\frac{m_f}{M_p} x^2$, then we automatically obtain $\frac{m_f}{M_p}  \phi^2 \ll \Lambda_{\text{eff}}^2$.
For chaotic inflation, which needs large $m_f$ and reaches transplanckian values of $ \phi $, this  implies a severe lower bound for $\Lambda_{\text{eff}}$, 
as explained in \cite{Dudas:2016eej}.

Second, one can consider the possibility of a dynamical $\Lambda_{\text{eff}}$, as we assumed in our illustrative model. Note that the derivation of the sgoldstino VEV and the EFT analysis would be
only mildly changed, since the \Kahler attractor potential reduces to the canonical shift symmetric one for small values of $\phi$ and $\alpha=1/3$. 
In particular, the conditions \eqref{regime_ine} should still be fulfilled. If $m_h \sim m_f$ the third condition in \eqref{regime_ine} would imply $\Lambda_{0} \gg M_{p}$ which is not possible. 
Said differently, the sgoldstino VEV would exceed the limit of the EFT validity if $m_h \sim m_f$, rendering the EFT description, with its truncation at quartic order in $S$, inconsistent.

A very careful different choice of the functions $f(\Phi)$, $h(\Phi)$ and $\Lambda_{\text{eff}}(\Phi)$ could possibly overcome these issues, however we consider them as a sufficient motivation to forgo chaotic inflation and prefer 
to work with the $\alpha$-attractor scenario.

\section{UV completion and mediation of SUSY breaking}\label{UVmodel}
We now consider a UV completion of the effective model whose aim is two-fold. We add extra fields which, upon integrating them out, produce the effective quartic term in $S$, and at the same time act as mediators of supersymmetry breaking to the standard model. As our goal is to achieve scales of SUSY breaking which are rather low, we set ourselves in a gauge mediation scenario \cite{Giudice:1998bp}, so that the extra fields can be taken to be the usual messengers, charged under the SM gauge group. 

As is customary in GMSB, it is better to actually take the messengers to be in representations of a grand unified group, $SU(5)$ being the minimal choice. Messengers come in vectorial representations, and the simplest option is to have one or more copies of ${\bf 5} \oplus \bar{\bf 5}$.

We thus introduce the fields $X$ and $Y$ in the  ${\bf 5}$, and $\tilde X$ and $\tilde Y$ in the $\bar{\bf 5}$, and we consider a 
generalized O'Raifeartaigh model which is equivalent to the effective IR description of the theory considered in \cite{Kitano:2006xg}.

In the rigid limit $M_p\to \infty$, we assume a canonical K\"ahler potential for all the fields (inflaton $\Phi$ included, which is indeed what one gets in this limit from \eqref{alphakahler} for $\alpha=1/3$), and a superpotential
\begin{equation}
W= - m_h \Phi^2 + f_0 S - \lambda_f S\Phi^2  +\lambda S X\tilde X
+(M+g_f \Phi) (X\tilde Y + Y \tilde X) + m_y Y\tilde Y\ .
\end{equation}
Let us comment on the various terms. If we set $\Phi=0$ and $m_y=0$, we have a complexified version of the usual O'Raifeartaigh model, with $S$ playing the role of the superfield with a non-vanishing F-term, a classical flat direction and the Goldstino as fermionic component. Its coupling to the messengers $\lambda$ can be taken to be typically $\mathcal{O}(1)$. Turning on the term with $m_y$ breaks R-symmetry and thus allows for gauge mediation to generate non-zero gaugino masses, 
as shown in 
\cite{Kitano:2006xg}. 

Bringing now the inflaton into the game, the crucial term is the one with $g_f$, which couples $\Phi$ to all the messengers and, more importantly, makes their mass inflaton-dependent. The last two terms involving $m_h$ and $\lambda_f=m_f/M_p$ will not be important for the physics of SUSY breaking. Note that, if we assume that the term with $m_h$ preserves R-symmetry, then both $g_f$ and $\lambda_f$ break it, and should be taken to be small. Indeed, $\lambda_f$ can be assumed to take its value between $10^ {-6}$ and $10^{-4}$  as in the previous section.\footnote{At this point, note that $W$ has also several terms that break the shift symmetry of the inflaton (i.e. the imaginary part of $\Phi$). The only one specific to the UV completion is the one proportional to $g_f$, which then cannot take too large values.}

Since $W$ breaks R-symmetry for generic couplings, we expect to have SUSY vacua.
However, if the R-symmetry breaking terms are small enough, we also expect the SUSY breaking vacuum near the origin to survive as a long-lived metastable state. We anticipate that the latter state will be obtained as follows: one sets all the messenger fields and the inflaton to zero, and then computes the effective potential for $S$. There is generically a local minimum near the origin 
%(the VEV of $S$ being small and  loop-suppressed) 
which becomes a global minimum if one takes $m_y, g_f, \lambda_f\to 0$. The SUSY vacua (i.e.~solutions to the F-term equations descending from $W$ above) occur either for parametrically large non-zero values of the messenger fields, i.e.~outside the domain of validity of the effective low-energy theory where the messengers have been integrated out, or for large sinflaton VEVs. It is indeed possible to check that the SUSY vacua are far enough from the inflationary trajectory in the $\Phi-S$ plane. 

\subsection{No tachyons in the messenger sector}
\label{UVtachions}
First, we discuss the conditions for which there are no tachyons in the messenger sector, along the inflationary trajectory, i.e. around zero values of $X$, $Y$, $\tilde X$ and $\tilde Y$. The potential, given by the sum of squared F-terms, is
\begin{align}
V=& |-2m_h \Phi - 2\lambda_f S\Phi+g_f(X\tilde Y + Y \tilde X)|^2 + |f_0 - \lambda_f \Phi^2 -\lambda X\tilde X|^2 \nn \\
& + |\lambda S X +(M+g_f \Phi)Y|^2 +   |\lambda S \tilde X +(M+g_f \Phi)\tilde Y|^2 \\
& + |(M+g_f \Phi) X + m_y Y|^2 + |(M+g_f \Phi) \tilde X + m_y\tilde Y|^2\ . \nn
\end{align}
Expanded to quadratic order in the messenger fields (but to any order in $S$ and $\Phi$), we get
\begin{align}
V=& -2 g_f\Phi^* (m_h+\lambda_f S^*)(X\tilde Y + Y \tilde X) -2 g_f\Phi (m_h+\lambda_f S)(X^*\tilde Y^* + Y^* \tilde X^*)  \nn \\
& -\lambda(f_0 - \lambda_f {\Phi^*}^2)  X\tilde X- \lambda(f_0 - \lambda_f \Phi^2)  X^*\tilde X^* \nn \\
& + \lambda^2 |S|^2(|X|^2+|\tilde X|^2) + |M+g_f \Phi|^2 ( |X|^2+|\tilde X|^2+|Y|^2+|\tilde Y|^2)+ m_y^2( |Y|^2+|\tilde Y|^2)\nn\\
& + \lambda S (M+g_f \Phi^*) (X Y^* + \tilde X \tilde Y^*) +  \lambda S^* (M+g_f \Phi) (X^* Y + \tilde X^* \tilde Y)\nn \\
& +  m_y(M+g_f \Phi) (X Y^* + \tilde X \tilde Y^*) + m_y (M+g_f \Phi^*)(X^* Y + \tilde X^* \tilde Y)\ . \label{vmessquad}
\end{align}
If we are to safely integrate out the messengers, we need to make sure that their squared mass matrix does not have negative eigenvalues, i.e. there are no tachyonic directions. When this is ensured, we will assume that the mass eigenvalues are dominated by the diagonal values (the third line in  \eqref{vmessquad}).\footnote{More specifically we can further assume that the diagonal terms are dominated by the term $|M+g_f \Phi|^2$. We can then distinguish two regimes, when $\Phi\gg M/g_f$ and when $\Phi\ll M/g_f$.} Thus in order to get a flavor of when tachyons could possibly arise, one can simply compare the off-diagonal terms to the diagonal ones. 

First of all, we must exclude tachyons near the origin, that is for values of $S$ and $\Phi$ subleading to any other scale. This is as in usual minimal GMSB, and we find 
\begin{equation}\label{notachyongmsb}
M^2\gtrsim |\lambda f_0|\ ,
\end{equation} 
where we have also assumed that $m_y$ is at most of the order of $M$ (this is necessary in order not to bring the SUSY vacua too close to the origin in the messenger directions). 

Going now to early times, at the beginning of the inflationary trajectory, we can assume $g_f \Phi \gg M$. In this regime the diagonal terms are all of the order of $g_f^2 \Phi^2$, while the off-diagonal terms are respectively of order $g_f \Phi (m_h+\lambda_f S)$, $\lambda \lambda_f \Phi^2$, $\lambda S g_f \Phi$ and $m_y g_f \Phi$. The new conditions are
\begin{equation}
g_f^2 \gtrsim \lambda \lambda_f\ ,\qquad 
  g_f \Phi\gtrsim \lambda S\ ,\qquad g_f \Phi\gtrsim m_h \ ,
\end{equation} 
where we have already simplified some redundant conditions by taking into account that   $m_y\lesssim M$ and $\lambda_f\ll \lambda$. 

Finally taking values of $\Phi$ which are such that $g_f\Phi \lesssim M$, the same rule of thumb that off-diagonal terms should be less than $M^2$ gives the only additional conditions
\begin{equation}
g_f \Phi \lesssim \frac{M^2}{m_h}\ , \qquad
\lambda S \lesssim M\ .
\end{equation}
The two conditions $g_f \Phi\gtrsim m_h$ and $g_f \Phi \lesssim M^2/m_h$ together imply that the inflationary trajectory does not cross a tachyonic region only if 
\begin{equation}
M\gtrsim m_h\ .
\end{equation}
Note that this condition implies that the mass of the inflaton after inflation lies within the regime of validity of the effective theory where the messengers have been integrated out.

All in all the no-tachyon constraints are
\be
\label{allinallUV}
M\gtrsim m_h\ , \qquad \lambda S \lesssim M+g_f \Phi \ , \qquad g_f^2 \gtrsim \lambda \lambda_f \ ,\qquad M^2\gtrsim |\lambda f_0|\ .
\ee
These are indeed the conditions we had been imposing on the EFT of the previous section, once the UV parameters are matched to the effective theory,
as we will see in the next subsection.

\subsection{Integrating out the messengers}
We now integrate out the messengers in order to make contact with the low-energy effective theory discussed in the previous section. Assuming that we are far away from the tachyonic regions, we can integrate out the messenger in a SUSY fashion, obtaining the one-loop K\"ahler potential directly \cite{Grisaru:1996ve,Intriligator:2006dd}. This can be further simplified by treating the terms proportional to  $\lambda S$ and $m_y$ as perturbations with respect to the dominant mass term proportional to $M+g_f \Phi$. 
The one-loop K\"ahler potential thus involves, when expanded in $S$, the following terms  
\bea
\label{1loopK}
K_\mathrm{1-loop}=
&&
-\frac{N_m \lambda^4(S S^{\dagger})^2}{12 (8 \pi^2)|M+g_f \Phi|^2  } 
-\frac{N_m \lambda^2}{2 (8 \pi^2)} S S^{\dagger}  \log [\frac{|M+g_f \Phi|^2}{M_p^2}] \\
&&
-\frac{N_m \lambda m_y}{2 (8 \pi^2)} \left( \frac{S^{\dagger} (M+g_f \Phi)^2+ h.c. }{|M+ g_f \Phi|^2} \right) +
\frac{N_m m_y \lambda^3}{12 (8 \pi^2)} S S^{\dagger} \left( \frac{S}{(M+g_f \Phi)^2}+ h.c. \right) 
\nonumber
\eea
where $N_m$ is the number of copies of the system composed of the $X,\tilde X, Y$ and $\tilde Y$ superfields. We actually have $N_m=5 N_5$, taking into account that the messengers come in representations $5$ and $\bar 5$.
Note that there are also one-loop terms dependent only on $S$ or only on $\Phi$, that we omitted since they just represent loop-suppressed corrections to the canonical terms.
The expression is expanded at first order in $m_y/(M+g_f \Phi)$.

The first term in \eqref{1loopK} corresponds to the higher dimensional correction that we considered in the EFT description (see Eq.~\eqref{alphakahler}) and is the one giving a mass to the sgoldstino.
The other terms are other one-loop corrections which are allowed by symmetries and indeed can be generically added to the effective \Kahler potential in \eqref{alphakahler}.
The second term in the first line is simply a sub-leading correction to the canonical kinetic term for the sgoldstino, which
depends on the inflaton $\varphi$ only at order $O(g_f^2)$.

The second line is suppressed by a factor of $m_y/M$ and as soon as $m_y < M$ it is not relevant for the inflaton trajectory.
However, these terms include the leading order operator that determines the decay of the inflaton in the sgoldstino, and are hence important later on in our analysis.
Moreover, they lead to a VEV for the sgoldstino also in the non-SUSY vacuum of the (rigid) theory at the end of inflation, which scales
as $\langle s \rangle \sim \frac{m_y}{\sqrt{2} \lambda}$. Note that for $m_y$ close to $M$, which will be needed in order to have sizable gluino masses, the sgoldstino VEV could become a relevant contribution to the inflaton mass which 
is now
$m_{  \varphi} \simeq 2 m_h + 2 \lambda_f \langle S \rangle$.
We will comment later about this fact and we will show that in the interesting region of the parameter space this contribution is always negligible so that we can simply take $m_{  \varphi} \simeq 2 m_h$, as in the discussion of section \ref{EFTanal}.

As a consistency check, we verified numerically on the benchmarks considered in the following analysis, that these extra \Kahler corrections play a negligible role in the determination of the inflaton trajectory in the $S-\Phi$ plane,
and the first term in \eqref{1loopK} is enough to capture the main features. 

Comparing \eqref{1loopK} with \eqref{alphakahler}, we see that we can identify 
\begin{equation}
\Lambda_0 + g \Phi = \frac{\sqrt6(4\pi)}{\sqrt{N_m}\lambda^2} (M+g_f\Phi)\ .
\end{equation}
As a consequence, the effective theory parameters $\Lambda_0$ and $g$ are related to the UV messenger theory parameters $M$ and $g_f$ by
\begin{equation}
\label{equation}
\Lambda_0 = \frac{\sqrt6(4\pi)}{\sqrt{N_m}\lambda^2} M\ , \qquad 
g  = \frac{\sqrt6(4\pi)}{\sqrt{N_m}\lambda^2} g_f\ .
\end{equation}
We see that, as it is usually the case, the UV scale of the effective theory, and its coupling $g$, are somewhat larger than the messenger masses, and their coupling to the inflaton, respectively. This has to be taken into account when comparing the inequalities keeping us within the validity of the effective theory, to the ones keeping us away from the tachyonic domain of the UV theory.
In particular,  if we use the matching \eqref{equation} to compare the inequalities that we have obtained in the EFT analysis in Section \ref{TheModel} (\Eq{easyconstraints} and \Eq{sgoldbis}) and the ones that we have obtained by demanding absence of tachyons in the UV theory in Section \ref{UVtachions} (\Eq{allinallUV}), we find that the former are typically weaker. 
Hence 
in the following we will consider the dynamics of the UV completed model and we will apply the conditions \eqref{allinallUV}.

\subsection{Low-energy spectrum and gauge mediation}
We now discuss the spectrum of the model, both during inflation and at the end of it. This will help setting the scale of some of the parameters of the model. From the inflationary sector we have the inflaton, together with its bosonic and fermionic partners. In the SUSY breaking sector, we have the gravitino and the sgoldstino. Eventually, we have the visible sector: we no longer consider the messengers, however their mass scale affects the visible sector soft masses through gauge mediation. 

Let us start with the inflationary sector. We concentrate here on the spectrum after inflation, which determines how much this sector could be relevant also to collider physics. Near the origin of $\Phi$, the SUSY mass of all its components is controlled by $m_h$. The SUSY breaking splittings are given by $\lambda_f f_0$, which is smaller than $m_h^2$ by virtue of the condition \eqref{notachion} to avoid tachyons in this sector.

In the $S$ sector, we can consider the mass of the gravitino and the mass of the sgoldstino. Note that the splitting between the latter two masses is all important for the viability of a sgoldstino-less description of inflation in supergravity. Indeed, we need to find the existence of a regime in which the gravitino is well below the scale of inflation (so that a supergravity description is justified) while the sgoldstino is above that scale, so that it makes sense to integrate it out (using a nilpotent superfield from the outset, for instance).

The mass of the gravitino is given by the standard SUGRA expression
\begin{equation}
m_{3/2}^2 = e^\frac{K}{M_p^2} \frac{|W|^2}{M_p^4} \simeq \left|\frac{h_0}{M_p}-m_h\frac{ \Phi^2}{M_p^2}\right|^2\ ,
\end{equation}
where we have used the fact that in all regimes $h(\Phi)\gg S f(\Phi)/M_p$. (In the large $\Phi$ regime, one uses  \eqref{vevS} to obtain $S\sim M_p \frac{ g^2 m_h}{m_f}$, under the only assumption $m_f\gg m_h$.) 
During inflation, the $\Phi$-dependent term will dominate, but given that $\Phi$ is Planckian at most, we will have $m_{3/2} \lesssim m_h$, indeed smaller than the scale of inflation which is determined by $m_f$.
At the end of inflation, the gravitino mass is given as usual by 
\begin{equation}\label{m32end}
m_{3/2}  \simeq \frac{f_0}{\sqrt 3 M_p}\ .
\end{equation}
In low scale SUSY breaking models, this will be the smallest scale. 

The sgoldstino mass is controlled by the quartic term in the K\"ahler potential \eqref{1loopK}. It gives a mass
\begin{equation}
m_s^2 = \frac{ N_m\lambda^4 |f_0 - \lambda_f \Phi^2|^2}{3(8\pi^2)|M+g_f\Phi|^2}\ .
\end{equation}
During inflation, i.e.~for large $\Phi$, we have
\begin{equation}
m_s \simeq \frac{\sqrt{ N_m}\lambda^2\lambda_f \Phi}{\sqrt 6 (2\pi)g_f} =2m_f\frac{\Phi}{gM_p}\ .
\end{equation}
Thus as long as $\Phi > g M_p$ the sgoldstino mass is larger than the scale of inflation, allowing to integrate it out (as in a nilpotent formulation).
However we also see that by no means it decouples entirely from the spectrum, its effective mass soon plunging below $m_f$. Indeed, at the end of inflation the sgoldstino mass is
\begin{equation}
m_s\simeq \frac{\sqrt{ N_m}\lambda^2 f_0}{\sqrt6(2\pi)M}\ .
\end{equation}
As we will see instantly, this is at a scale just an order of magnitude larger than the soft masses of the visible sector.

Assuming gauge mediation of SUSY breaking to be exclusively operated through the messengers which also couple to the inflaton, the soft masses of the visible sector can be completely determined. We note that the expressions are complicated by the presence of several parameters, such as $m_y$ which is necessary to obtain non-zero gaugino masses \cite{Kitano:2006xg}. 
The gaugino and sfermion masses scale similarly to the sgoldstino mass, where however the loop suppression is due to SM gauge couplings, and where the gaugino suffers from an extra suppression in 
power of $m_y/M$ because of R-symmetry
\begin{equation}
m_\mathrm{sfermions}^2 \simeq \sum_i \frac{2 N_5 C_2^i g_{SM (i)}^4 \lambda^2 |f_0 - \lambda_f \Phi^2|^2}{(4\pi)^4|M+g_f\Phi|^2}\ , \qquad m^{(i)}_\mathrm{gaugino} \simeq \frac{N_5 g_{SM(i)}^2 \lambda m_y |f_0 - \lambda_f \Phi^2|}{(4\pi)^2|M+g_f\Phi|^2}\ .
\end{equation}
In particular the sfermion masses are quite large during inflation, but eventually reduce to the usual value \cite{Kitano:2006xg} after settling in the vacuum:\footnote{We neglect RG running effects.}
\begin{equation}
\label{softmasses}
m_\mathrm{sfermions}^2 \simeq \sum_i \frac{2N_5 C_2^i  g_{SM(i)}^4 \lambda^2 f_0^2 }{(4\pi)^4 M^2}\ , \qquad m^{(i)}_\mathrm{gaugino} \simeq \frac{N_5 g_{SM(i)}^2 \lambda m_y  f_0 }{(4\pi)^2 M^2 }\ ,
\end{equation}
where $C_2^i$ are the quadratic Casimir of the sfermions and $i$ runs over the SM gauge groups.
These are the values that we will use in the phenomenological analysis in Section \ref{Pheno}.

\subsection{Analysis of the allowed parameter space}
We can now combine all the consistency conditions in order to identify the allowed regions of parameter space and determine what are the typical mass scales of the relevant
particles entering into the model, i.e.~the inflaton, the sgoldstino, and the superpartners (we will consider as reference the gluino).

The model depends on many parameters, but eventually only few of them determine a qualitative difference in the physics outcome.
We summarize here the relevant constraints on the parameters that we have encountered along our analysis:
\begin{itemize}
\item No tachyon condition in the UV model, arising from the analysis of the messenger mass matrix:
\be
\label{tachyons1}
M\gtrsim m_h\ , \qquad \lambda S \lesssim M+g_f \Phi \ , \qquad g_f^2 \gtrsim \lambda \frac{m_f}{M_p} \ ,\qquad M^2\gtrsim |\lambda f_0|\ .
\ee
The condition on the sgoldstino VEV along the entire trajectory can be analyzed 
as we did for the EFT in \eqref{regime_ine} (see Appendix \ref{regimes}), yielding
\be \label{uvcondtraj}
\lambda S \lesssim M+g_f \Phi \quad \Rightarrow \quad \left\{ f_0  \lesssim \frac{m_f}{ \kappa M_p} M^2 \, , \, \frac{M^2}{\kappa M_p^2}\lesssim\frac{m_h}{m_f}   \,, \, \frac{M}{M_p} \gtrsim \lambda  \frac{m_h}{m_f}
\right\} \ ,
\ee
where $\kappa=\frac{N_m \lambda^4}{12 (8 \pi^2)} $. In practice, these conditions can be circumvented taking a strongly coupled hidden and messenger sector, for which no perturbative analysis can be done, leaving us with the EFT treatment of the previous section.

\item No tachyons in the effective theory, i.e.~the inflaton sector:
\be
\label{tachyons2}
m_h^2 > \frac{3 \alpha f_0 m_f}{2 M_p} \ .
\ee
\end{itemize}

It is not straightforward to extract from this set of inequalities what is the allowed volume in the parameter space. So, in order to investigate the allowed region, we can proceed by fixing some parameters to typical values and plot the resulting
region.
Once we fix the dimensionless quantities $(\lambda,g_f,N_5,\frac{m_y}{M})$ and we take $m_f \sim 10^{-5} M_p$ as suggested by the analysis in Figure \ref{fig:alphaObs}, we are left with only 3 independent parameters, i.e.~$\{ f_0, M, m_h \}$. We can trade two of these parameters with physical masses to conclude that our parameter space is a region in the three dimensional space spanned by $\{m_{\lambda}, m_{3/2}, m_h\}$, where
we indicated with $m_{\lambda}$ the gluino mass.

\begin{figure}[t]
\begin{center}
\includegraphics[width=0.7\linewidth]{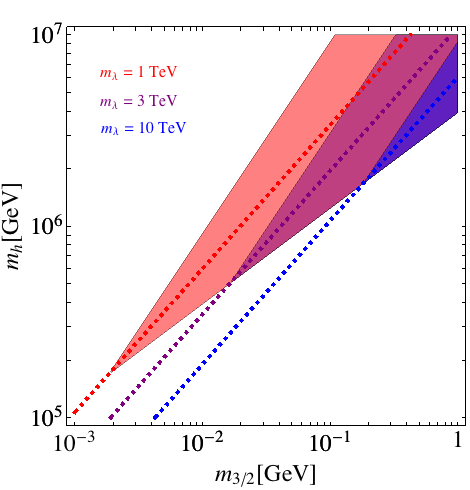}
\caption{\label{fig:parameterspace} Allowed region of parameter space in the $\{m_{3/2}, m_h\}$ plane by fixing $m_{\lambda} = 1,3,10$ TeV respectively for the light red, purple, blue region.
The dashed lines represent the line defined in equation \eqref{middleline} for the different gluino masses.
The other parameters are fixed as $\{\frac{m_y}{M}=0.15, \lambda=0.25, N_5=3, g_f=2 \times 10^{-3}, \alpha=1/3, m_f =7.3 \times 10^{-6}M_p \}$.}
\end{center}
\end{figure}

The allowed region of parameter space can then be easily displayed in the $\{m_{3/2}, m_h\}$ plane by fixing $m_{\lambda}$ to some phenomenologically interesting value, as we do in Figure \ref{fig:parameterspace}.
We considered as upper limit for the gravitino mass the value of $1$ GeV since we want to focus on the case where the gauge mediated contributions to the soft masses dominate the gravity contributions.%
\footnote{In producing the plot, we also imposed $M \leq 10^{15}$GeV.}

As we can observe, the allowed region in the $\{m_{3/2}, m_h\}$ plane gets smaller as we increase the gluino mass, 
disappearing completely (in the $m_{3/2} \leq 1$ GeV region) for $m_{\lambda} \geq 20$ TeV on the selected benchmark.
Note that the two boundaries are set by the two conditions (where we have reinstated the exact numerical prefactor for added precision):
\bea
\label{upperline}
&&\frac{M}{M_p} > \lambda \sqrt2 \frac{m_h}{m_f} \qquad \Longrightarrow \qquad \text{Upper border}\ , \\
\label{lowerline}
&&m_h^2 > \frac{3 \alpha f_0 m_f}{2 M_p}  \qquad \Longrightarrow \qquad \text{Lower border}\ .
\eea
We can saturate these inequalities to find the expressions $m_h^{(up)}$ and  $m_h^{(low)}$ which determine the upper and lower lines of the triangle shaped allowed regions
in Figure \ref{fig:parameterspace}.

It is instructive to consider the tip of the triangles, giving the lowest allowed values for the gravitino mass (and the inflaton mass), for a given gluino mass. It is simply obtained by saturating the two inequalities above. Using \eqref{m32end} and \eqref{softmasses} we get:
\begin{equation}
m_{3/2}^\mathrm{tip} = \sqrt3 \alpha \left(\frac{(4\pi)^2}{N_5 g_{SM}^2}\right)^2 \left(\frac{M}{m_y}\right)^2 \frac{m_\lambda^2}{m_f}\ .
\end{equation}
This is to be confronted with the lowest gravitino mass that one could obtain in our gauge mediated model, considered on its own, given by the bound \eqref{notachyongmsb}:
\begin{equation}
m_{3/2}^\mathrm{lowest,GMSB} = \frac{1}{\sqrt3\lambda} \left(\frac{(4\pi)^2}{N_5 g_{SM}^2}\right)^2 \left(\frac{M}{m_y}\right)^2 \frac{m_\lambda^2}{M_p}\ .
\end{equation}
We thus see that 
\begin{equation}
m_{3/2}^\mathrm{tip} = 3 \alpha \lambda \frac{M_p}{m_f}m_{3/2}^\mathrm{lowest,GMSB}\ ,
\end{equation}
that is we are roughly 5 orders of magnitude above the lowest gravitino masses generically allowed by GMSB.\footnote{We can actually trace back this bound to one of the conditions in \eqref{uvcondtraj}, specifically the one giving an upper bound to $f_0/M^2$ proportional to $m_f/M_p$.} In other words, the scale of supersymmetry breaking cannot be as low as we could hope in a GMSB scenario. Putting numbers, and using the benchmark point of Figure \ref{fig:parameterspace}, we get
\begin{equation}
m_{3/2}^\mathrm{tip} \simeq 2\ \mathrm{MeV} \left(\frac{m_\lambda}{1\ \mathrm{TeV}}\right)^2\ .
\end{equation}
Thus we see that, for reasonable gluino masses, the gravitino cannot be lighter than a few MeV. This translates into a lowest supersymmetry breaking scale of the order of $\sqrt{f_0}\sim 10^8$~GeV,
showing that for this class of weakly coupled models the scale of SUSY breaking cannot be arbitrarily small and compatible with sgoldstino-less inflation.

\section{Phenomenological analysis}\label{Pheno}

In order to simplify our analysis, from now on we will focus on the dashed lines in the middle of the triangles in Figure \ref{fig:parameterspace}, which is the average mean defined as
\be
\label{middleline}
m_{h}^* = \sqrt{m_h^{(up)} m_h^{(low)}}\ ,
\ee
which determines $m_h$ for a given value of the other parameters.
In this way we are reduced to a two dimensional parameter space spanned by $\{ m_{3/2},m_{\lambda} \}$, in which we will present our phenomenological analysis.

Given that the model we have considered in the previous section includes predictions both for cosmology as well as for particle physics, we can constrain the parameter space using 
 inflation observables, considerations about the reheating temperature, gravitino dark matter abundance, Big Bang nucleosynthesis (BBN), as well as LHC constraints.
We discuss all these aspects in the next subsections under some simplifying assumptions.
In particular, 
we do not take into account the extended Higgs sector and its possible effects on the phenomenology. 
Moreover, we consider the neutralino NLSP to be predominantly Bino, assuming that the $\mu$ parameter is such that the Higgsino is significantly heavier,
and
consistent with the estimates of the soft terms done previously \eqref{softmasses}. 
At the end of the section, 
we will comment on how our analysis would be affected if instead the NLSP neutralino is a mixture of Bino-Higgsino.

We will present all the phenomenological characterization in the $\{ m_{3/2},m_{\lambda} \}$ plane, restricting to the line \eqref{middleline}
as just mentioned.
We verified that exploring other areas of the allowed region in Figure \ref{fig:parameterspace} does not change qualitatively our conclusions.
 
 Our investigation will show that, once we  consider all bounds together, the remaining allowed parameter space gets significantly reduced.
 In particular our results highlight the complementarity of the different phenomenological constraints, suggesting that 
 a broad approach to inflation models, including analysis of the reheating epoch as well as the connection to particle physics,
 is needed in order to extract robust conclusions and predictions.

\subsection{Reheating temperature and $n_s$}

Since in our model the couplings between the inflaton and the MSSM particles are well defined (up to the Higgs sector, that we do not specify), 
we can estimate the inflaton decay modes and the reheating temperature.
This is relevant since the inflaton decays via messenger loops to MSSM particles and thus the reheating temperature will be low. A low reheating temperature corresponds to a relatively small number of e-folds during inflation, hence a sizeable shift in $n_s$ %and $r$
compared to the usual estimate of 60 e-folds, as in  \cite{Ellis:2015pla,Ueno:2016dim}. 
In addition, the gravitino problem is simpler to solve with a low reheating temperature
\cite{deGouvea:1997afu,Rychkov:2007uq}.

The reheating temperature can be found from the energy density at which the Hubble rate ($H_{re}$) equals the decay width of the inflaton ($\Gamma_\phi$) \cite{Kofman:1994rk}
\be
	\rho_{re}=3H_{re}^2M_p^2=3\Gamma_\phi^2M_p^2=\rho_{end}e^{-3N_{re}(1+\bar{w}_{re})}=\frac{\pi^2 g_*}{30}T_{re}^4,
\ee
where $H_{re}$ is the Hubble rate at the end of reheating, $g_*\sim 220$ is the number of relativistic degrees of freedom at reheating, $\bar w_{re}$ is the average equation of state during reheating and $\rho_{end}$ is the energy density at the end of inflation. The third equality defines the number of e-folds during reheating ($N_{re}$) and the last equality the temperature $T_{re}$. Though the thermalization after the inflaton decay might take time and result in a lower reheating temperature, we will in the following stick to this upper bound. Finally, since our potential is quadratic around the minimum, $\bar{w}_{re}=0$ \cite{Turner:1983he,Munoz:2014eqa}. 

In the CMB we observe the modes at the pivot scale $k=0.002 \text{ Mpc}^{-1}$. During inflation this scale was the horizon size ($a_k H_k$), so we can write
 \cite{Liddle:2003as}
\be
\label{eq_reh_as}
	\frac{k}{a_0H_0}=\frac{a_k}{a_{end}}\frac{a_{end}}{a_{re}}\frac{a_{re}}{a_{eq}}\frac{a_{eq} H_{eq}}{a_0 H_0}\frac{H_k}{H_{eq}},
\ee
where the subscripts `$end$', `$eq$' and `0' correspond to the end of inflation, matter-radiation equality and the current time respectively. Taking the logarithm of Eq.~(\ref{eq_reh_as}) gives us, after some manipulations, the number of e-folds during inflation as a function of the number of e-folds during reheating and other known constants \cite{Munoz:2014eqa}
\be 
\label{eq_reh_Nstar}
	N=\frac{1}{4} N_{re}-\Log{\frac{k}{a_0 T_0}}-\frac{1}{4}\Log{\frac{30}{g_*\pi^2}}-\frac{1}{3}\Log{\frac{11 g_*}{43}}-\frac{1}{4}\Log{\frac{3}{2}\frac{V_{end}}{M_p^4}}+\frac{1}{2}\Log{\frac{\pi^2 r A_s}{2}}.
\ee
Note that this is an implicit equation since $r$ depends on $N$
as shown in \Eq{eq_nsrN}. This relation can be solved for the reheating temperature, resulting in 
\begin{equation}
\label{eq_Treh_N}
	T_{re}=\frac{495\sqrt{3}}{43\pi^2\sqrt{2A_s\alpha}}M_p\left(\frac{k}{a_0 T_0}\right)^3 \left(\frac{4+\sqrt{16+3\alpha}-\sqrt{3\alpha}}{4+\sqrt{16+3\alpha}+\sqrt{3\alpha}}\right)^{4}\left(N+{\cal O}\left(N^0\right)\right)e^{3N},
\end{equation}
where again we used the $1/N$ expansion. A more careful analysis, keeping higher orders in $N$ showed that in our model corrections to this equation are of the order of a percent. However, due to the exponential behaviour on the number of e-folds, a small deviation in this quantity changes the reheating temperature considerably.

\begin{figure}[t]
\begin{center}
\includegraphics[width=0.32\linewidth]{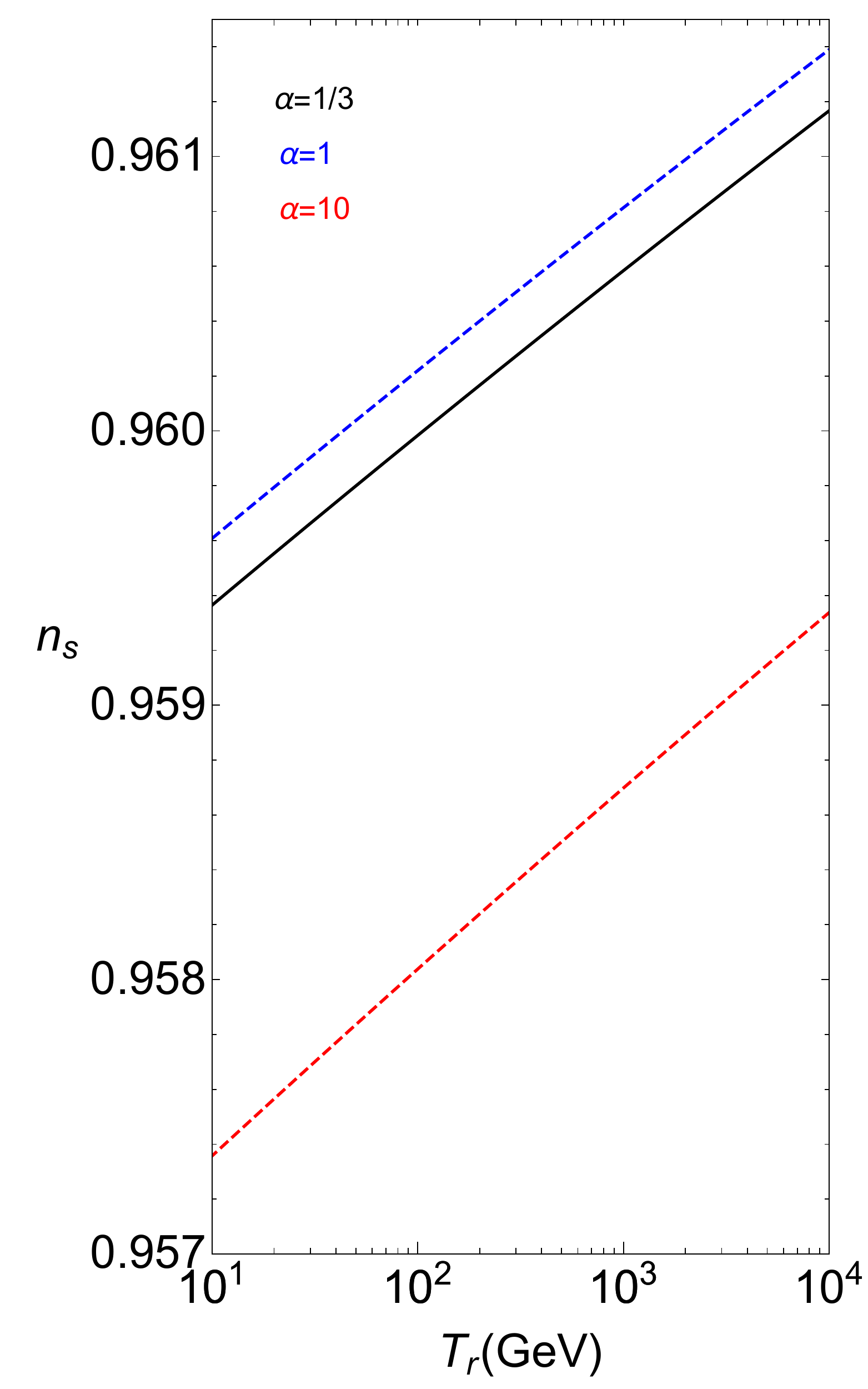}
\includegraphics[width=0.32\linewidth]{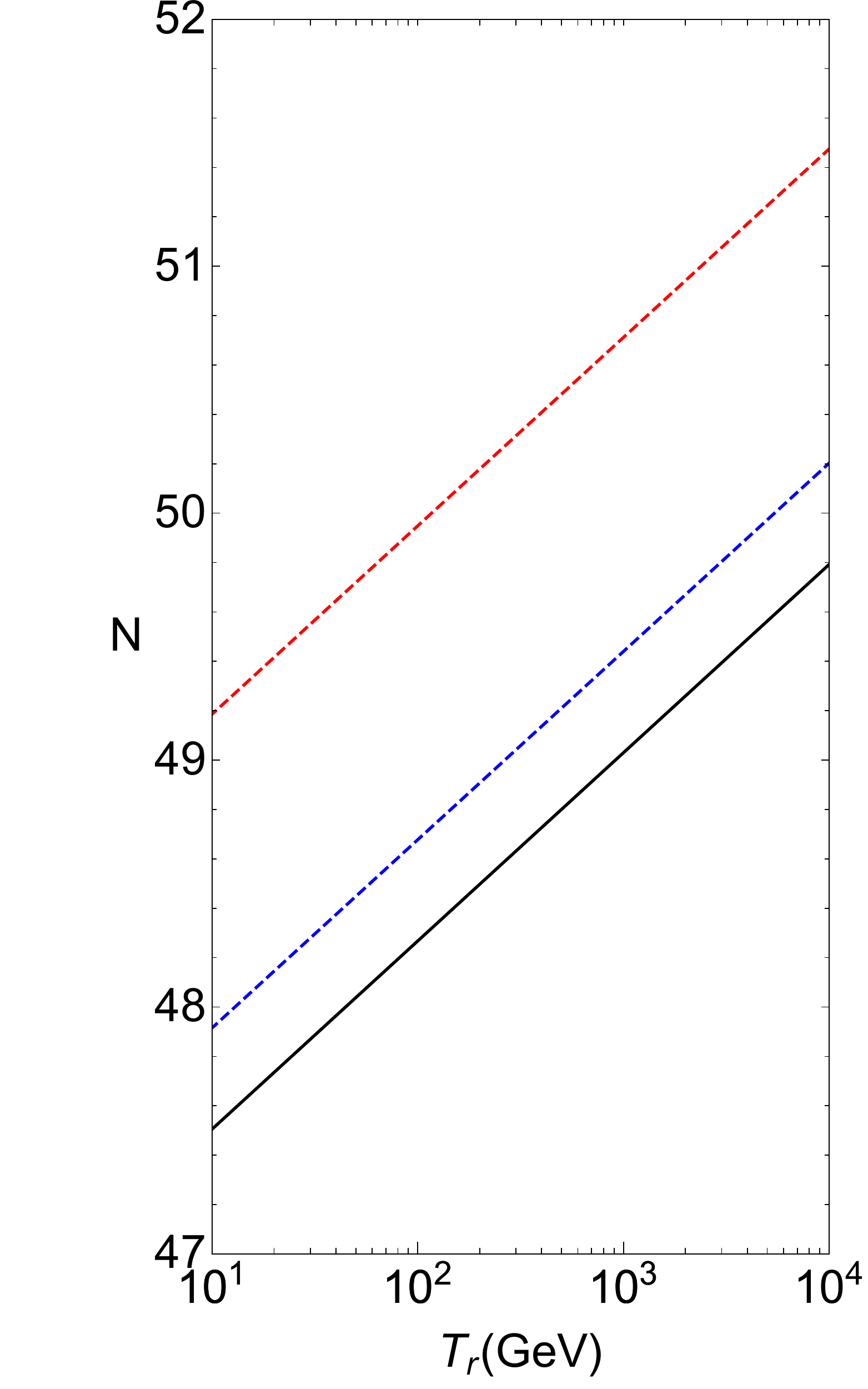}
\includegraphics[width=0.32\linewidth]{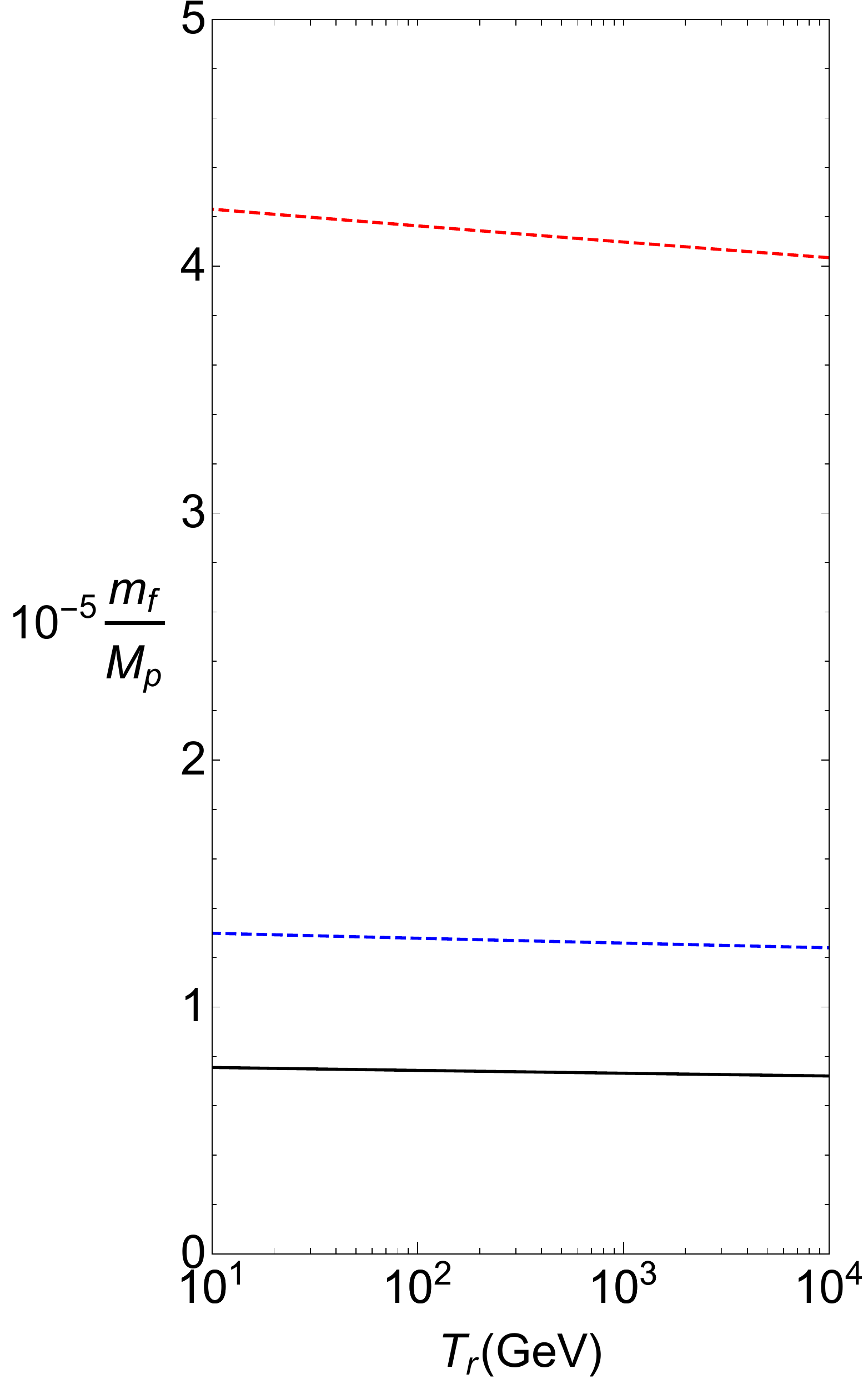}
\caption{\label{fig:Trns} 
From left to right, the dependence on the reheating temperature of $n_s$, $N$ and $m_f$ for different values of $\alpha$. The straight black line represents our scenario with $\alpha=1/3$, while the blue and red dotted lines represent resp.~$\alpha={1,10}$.}
\end{center}
\end{figure}

In our model the reheating temperature can be computed from the inflaton decay width, and we  
use   \Eq{eq_reh_Nstar} to extract the expected number of e-folds.
Then we plug this result in \Eq{eq_nsrN} and \Eq{mffix} to obtain precise predictions for $n_s$, $r$ and $m_f$ as a function of the reheating temperature. The results of this procedure are plotted in \Fig{fig:Trns} for $N$, $n_s$ and $m_f$. The tensor to scalar ratio, being $1/N^2$ suppressed, is roughly $0.002$, far below the current experimental constraints.

We now discuss the most relevant decay modes of the inflaton and estimate its decay width to obtain the expected reheating temperature.
The mass of the inflaton is taken to be $ m_{ \varphi} = 2 m_h$, neglecting possible contributions from the sgoldstino VEV. 
We will verify that this is consistent on the considered region of the parameter space.

In the complete model there is no tree-level coupling between the inflaton and MSSM fields and the decay channels are loop suppressed.
The leading decay mode arises from the following operator generated at one-loop (together with the operator responsible for the gluino masses)
\be\label{phiWW}
W \supset \int d^2\theta  \frac{\alpha_s N_5}{4 \pi}   \frac{ g_f}{M}  \Phi  \, \mathcal{W}_{\alpha}\mathcal{W}^{\alpha}\,,
\ee
and it gives the following decay modes into gluons and gluinos%
\be
\label{phigg}
	\Gamma_\phi\approx\Gamma_{\phi gg}+\Gamma_{\phi \tilde g \tilde g} =
\frac{ \alpha_s^2 N_5^2 g_f^2} {\pi^3 M^2}  m_h^3 \ .
\ee
These are the dominant decay modes of the inflaton\footnote{The ones into sfermions are loop suppressed and the ones in other gauge bosons/gauginos are coupling
suppressed.} and
we approximate with this value the total decay width of the inflaton.
Because of the one-loop suppression of this decay width the reheating temperature is rather low in our parameter space, 
of the order of the TeV scale,
\be
T_{re} \simeq 900 \text{ GeV} \left (\frac{m_h}{200 \text{ TeV}} \right)^{3/2} \left(\frac{m_{\lambda}}{1 \text{ TeV}} \right)  \left(\frac{20 \text{ MeV}}{m_{3/2}} \right)
\ee
where the numerical values of the parameters has been fixed as in Figure \ref{fig:parameterspace}.
Then, from Figure \ref{fig:Trns} we see that for $\alpha=1/3$ the number of e-folds is about $48-49$, and hence that $n_s$ is rather small. 
We also note that the value of $m_f$ is around $ 8 \times 10^{-6}M_p$, as the benchmark point we have chosen in the previous section.

\subsection{Gravitino abundance}

In this model the gravitino is the LSP and can be a viable dark matter candidate
if $\Omega_{3/2} h^2 \simeq \Omega_{DM} h^2 = 0.12$ 
(see e.g. \cite{Benakli:2017whb,Dudas:2017rpa,Co:2016fln,Co:2017orl} for recent developments on gravitino dark matter).
Neglecting possible dilution factors, we conservatively demand that the gravitino abundance does not overclose the universe by imposing
$\Omega_{3/2} h^2 \leq \Omega_{DM} h^2 $. 
Gravitino relics can be obtained with several mechanisms: i) thermal production;  ii) freeze-in production through the superparticle decays; iii) production through inflaton decays; iv) decay of the NLSP, that here is assumed to be the Bino.

In the following we study these production mechanisms and we find the constraints they impose on the parameter space of the model:

\begin{itemize}
\item[i)] The reheating temperature for our range of parameters is at most of  $O(\text{TeV})$, with gravitino at least of MeV mass, and hence the thermal production of gravitino
is not significant \cite{Moroi:1993mb,deGouvea:1997afu,Rychkov:2007uq}.
We verified this explicitly in our numerical analysis.

\item[ii)]
The freeze-in scenario for gravitino dark matter has been proposed as a mechanism to obtain the correct gravitino relic abundance 
through the decay of the MSSM superparticles \cite{Moroi:1993mb,Hall:2009bx,Cheung:2011nn,Hall:2013uga}.
In our case the abundance of the heavy superparticles is suppressed by a Boltzmann factor, due to the low reheating temperature,\footnote{In the period between the end of inflation and before the radiation dominated era, the temperature can be higher, and hence also the freeze-in contribution. However, the produced gravitinos get diluted before radiation starts dominating \cite{Giudice:2000ex,Monteux:2015qqa,Co:2015pka}. As a crude approximation, we neglect these two compensating effects.} and hence we have to use the formula \cite{Hall:2009bx}
\begin{equation}
	Y_{\text{freeze-in}}=\sum_{\tilde{X}} \frac{45 \sqrt{90} g_{\tilde{X}}}{4\pi^5 g_*^{3/2}} \frac{\Gamma_{\tilde{X}}M_p}{m_{\tilde{X}}^2}\int_{x_{min}}^{x_{max}}K_1(x)x^3 \d{x},
\end{equation}
where the sum runs over all the superparticles $\tilde X$,  $g_{\tilde{X}}$ is the number of internal degrees of freedom of species $\tilde{X}$, $\Gamma_{\tilde{X}}$ is the partial decay width of $\tilde{X}$ into gravitinos,  $K_1$ is the first Bessel function and $x=m_{\tilde{X}}/T$ (where we suppressed the index $\tilde{X}$). The superparticle masses are much heavier than the current temperature of the universe, so we set $x_{max}=\infty$, however we cannot ignore the low reheating temperature, so $x_{min}=m_{\tilde{X}}/T_{re}$.  We numerically perform this integral (though it can be done  analytically), and demand this abundance to not exceed the dark matter one.

\item[iii)] The decay of the inflaton into gravitinos, or into supersymmetric particles eventually decaying into gravitinos, 
can give a large contribution to the gravitino relic abundance. In these processes the goldstino component of the gravitino is the one setting the relevant interactions.

The inflaton can decay directly into goldstinos or sgoldstinos through the interactions induced by the second line of \eqref{1loopK}.
The second term in the second line in equation \eqref{1loopK} determines the direct decay into goldstinos as
\be
\label{phitoGG}
\Gamma_{\varphi \to GG } =\frac{1}{4 \pi} \left( \frac{N_m  g_f\lambda^3 m_y}{48 \pi^2 M^3 \sqrt{2} } f_0  \right)^2  m_{\varphi}\ ,
\ee
where $ m_{\varphi}=2 m_h$ is the mass of the inflaton.
This decay is very much suppressed compared to \eqref{phigg} but can nevertheless lead to an overabundance of gravitinos.
Moreover, another relevant channel is the decay into sgoldstino, since in our parameter space we have $ m_{\varphi} > m_s$.
This decay mode has the same partial width
as the one above, hence it can be important since the sgoldstino will decay mainly into goldstinos.

We estimate the abundance of gravitinos arising from these processes, assuming conservatively that the sgoldstino decays with 100\% BR into two  
gravitinos,
as
\be
\label{infl_dec_32}
Y_{3/2}^{\text{decay}} \simeq  \frac{3 T_{re}}{4  m_{\varphi}} \left( 2 \text{BR}_{\varphi \to GG } + 4 \text{BR}_{\varphi \to s \sigma} \right) \ ,
\ee
where we included multiplicity factors.
We impose that such abundance does not exceed the dark matter abundance.

Furthermore, decays of the inflaton into MSSM particles and eventually into the NLSP have sizable branching ratios. However, if the NLSP is in thermal equilibrium, its abundance will be set by the thermal bath dynamics. 
In order for the NSLP to be in thermal equilibrium we demand
that the reheating temperature is larger than the NLSP mass, i.e.  the Bino mass
\be
T_{re} \gtrsim m_{\tilde B}  \ .
\ee

\item [iv)]
The abundance of the Bino in the case in which the other sparticles are very heavy has been computed in \cite{Drees:1992am,Wells:1998ci,Wells:1997ag} and it depends
on the slepton masses $m_{\tilde{l}}$, which we estimate in \eqref{softmasses}, and reads
\be
\Omega_{\tilde{B}}h^2 = \frac{\left(m_{\tilde{B}}^2+m_{\tilde{l}}^2\right)^4}{(460 \text{GeV})^2 \sqrt{g_*} m_{\tilde{B}}^2 \left(m_{\tilde{B}}^4+m_{\tilde{l}}^4\right)}\ .
\ee
The abundance of the Bino will set the gravitino abundance obtained via Bino decay as
\be
\Omega_{3/2}^{\tilde B} h^2 =\frac{m_{3/2}}{m_{\tilde B}} \Omega_{\tilde B} h^2\ .
\ee
In our analysis we will demand that $\Omega_{3/2}^{\tilde B} h^2 \lesssim \Omega_{DM} h^2$.

\end{itemize}

 The gravitinos produced by any of the processes aforementioned could potentially carry a large energy (at most $m_{\varphi}/2$). A too high free-streaming length of the gravitino could thus destroy small scale structures and is experimentally constrained \cite{Boyarsky:2008xj}. Since in our model the inflaton mass is very low (compared to other inflation models), we find, using results of \cite{Takahashi:2007tz} that this does not constrain our parameter space.

\subsection{BBN}
Late decays of the NLSP into gravitinos can spoil BBN if the decay time is larger than $\sim0.1$s  \cite{Kawasaki:2004qu,Kawasaki:2008qe}.
This poses an absolute lower bound in the $(m_{3/2},m_{\lambda})$ plane given by the requirement
\be
\label{BBN_bound}
\tau_{\tilde B} \simeq
\Gamma_{\tilde B \to \gamma/Z +G}^{-1} \simeq \frac{48 \pi m_{3/2}^2 M_{p}^2}{m_{\tilde B}^5} \simeq  0.1\ \text{sec} \left( \frac{m_{3/2}}{10 \, \text{MeV}} \right)^{2} \left( \frac{225 \,\text{GeV}}{m_{\tilde B}} \right)^5 \leq 0.1\ \text{sec}\ .
\ee

\subsection{Combination of the cosmological and LHC constraints}
We can now combine all the constraints listed above in one single plot which highlights the viable region in the parameter space of the model in the 
$(m_{3/2},m_{\lambda})$ plane. As mentioned, we restrict to the dashed line in Figure \ref{fig:parameterspace} for definitess (cfr. \eqref{middleline}).

\begin{figure}[t]
\begin{center}
\hspace{-1cm}\includegraphics[width=0.7\linewidth]{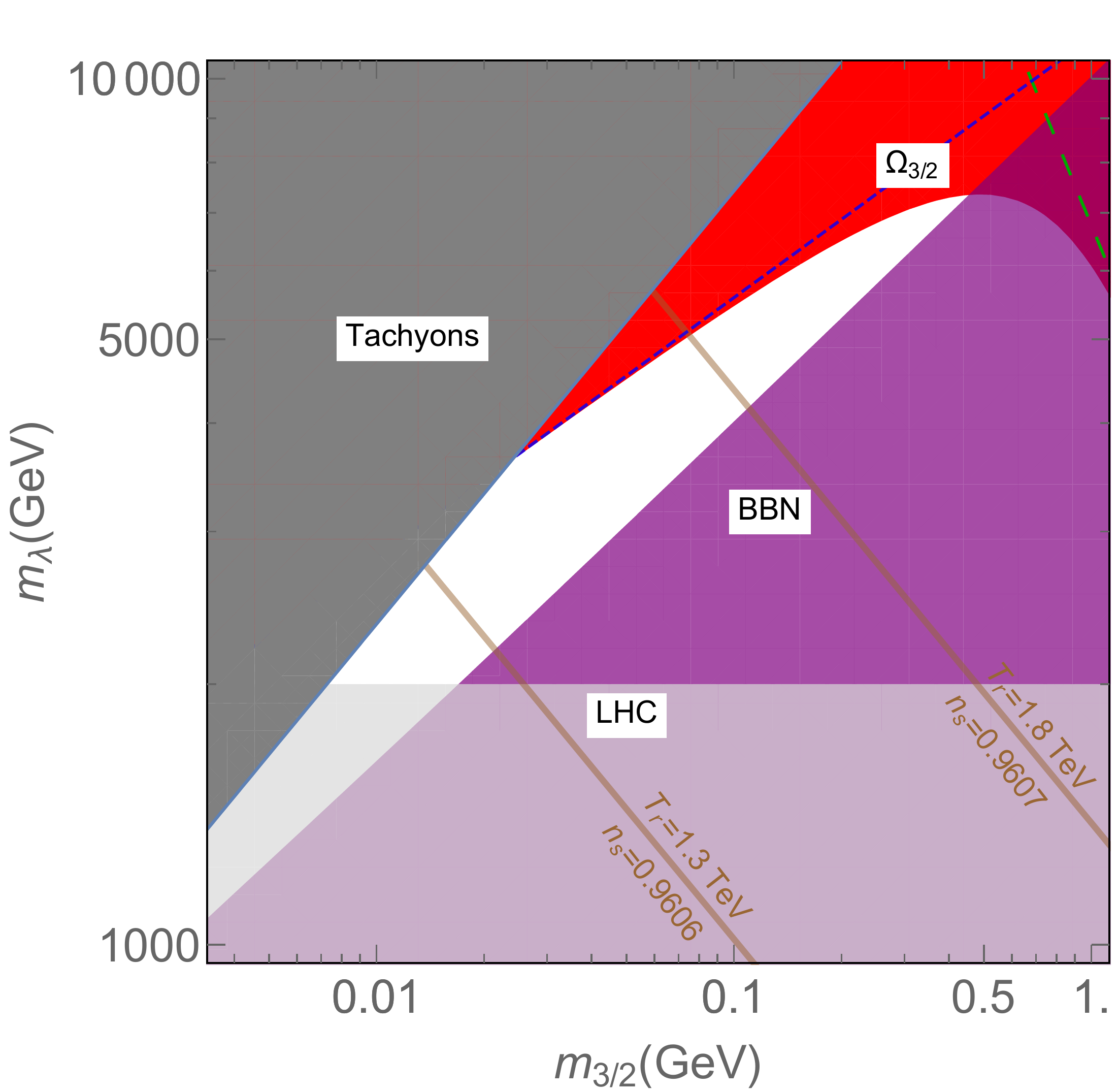}
\caption{\label{fig:money} Phenomenological constraints on the parameter space in the $\{m_{3/2}, m_{\lambda}\}$ plane by fixing $m_h$ with the relation \eqref{middleline}. 
The other parameters are fixed to the same numerical values as in Figure \ref{fig:parameterspace}.
The white region is the one satisfying all the constraints explained in the
text. The black region is excluded because of tachyon conditions. 
The red and purple regions are not viable because of phenomenological constraints on gravitino abundance or BBN, respectively. 
In the light blue region the gluino mass is lower than $2$ TeV, the current LHC bound \cite{ATLAS:2017cjl,CMS:2017gzh}.
The contours denote the spectral tilt $n_s$ and the reheating temperature.}
\end{center}
\end{figure}

In Figure \ref{fig:money} we show our results for a representative choice of the numerical parameters, which however does not influence the qualitative features of the 
conclusions.\footnote{One can verify numerically that in the explored region the inflaton mass is always given predominantly by $2 m_h$, corroborating the previous 
statements.}
The black region is excluded because of tachyons in the spectrum, and corresponds in the plot of Figure \ref{fig:parameterspace} to going beyond the tip of the allowed conical region along the dashed lines. 
The purple region is excluded because of the BBN constraint on the Bino decay time in \eqref{BBN_bound}.
The red region is excluded since the gravitino abundance exceeds the one expected for dark matter because of inflaton decay (above the dashed blue line), because of Bino decay (the region on the right of the
dashed green line), or because of a combination of these two mechanisms.
The freeze-in contribution to gravitino abundance, instead, turns out not to be relevant on the explored parameter space.

The allowed region passing all the constraints is the white spiky region.
On the border between the white and the red regions the gravitino has the correct relic abundance to be the dark matter of the universe. 
In the rest of the white region dark matter can be constituted by 
some other particle in the hidden sector, e.g.~as in \cite{Dimopoulos:1996gy}.

We note that the gluino mass is bounded from above in the allowed region, and should be smaller than around $7$ TeV for the choice of parameters in the plot.
In the blue region the gluino mass is below $2$ TeV, and hence excluded by the LHC 
bound.\footnote{This estimate is very conservative since the actual bound depends on the decay mode of the gluino and on the rest of the SUSY spectrum.} 
Similarly, also the gravitino mass has both a lower and an upper bound, 
spanning between a few MeV and half a GeV. In particular, it is interesting to see that the upper bounds on $m_\lambda$ and $m_{3/2}$ are generated by a combination of the constraints from gravitino overabundance and BBN.

In Figure \ref{fig:money} we also show the contours of the inflation observable $n_s$, obtained after computing the reheating temperature and using the results of Figure \ref{fig:Trns}.
Since a larger $n_s$ (hence larger reheating temperature) is favoured by the Planck data, it is interesting that both Planck and the LHC prefer the same region of parameter space, with
larger gluino and gravitino mass.
This is clearly a specific feature of the model considered, but it is appealing that two experiments probing completely different physics can provide indications on the same BSM theory. 

Our results should be considered as a preliminary analysis of an illustrative model which combines inflation and particle physics predictions, and it
emphasizes the interplay between these two sets of physical requirements in shaping viable scenarios.

 \paragraph{Further remarks}
There are several aspects that we did not consider in our analysis and could be improved.
First, in our model besides the inflaton there is another scalar field, the sgoldstino, that could potentially dominate the energy density of the universe at some stage of the
cosmological evolution. This would be problematic given that the sgoldstino decays significantly into gravitinos, eventually leading to an overabundance of the latter.
However we argue that this should not happen in the parameter space we explored for the following reasons, leaving a detailed analysis to future work.
The scenario is very similar to curvaton models \cite{Linde:1996gt,Bartolo:2002vf,Vennin:2015vfa}, with the sgoldstino playing the role of the curvaton.
First, as one can check explicitly, the decay width of the sgoldstino is of similar order of magnitude of the decay width of the inflaton
for our numerical values used for Figure \ref{fig:money}.
Moreover, the field excursion of the sgoldstino VEV is much smaller than the inflaton one, being constrained to be smaller than $M+g_f \phi$ to avoid tachyons, and being concretely always smaller than $M$ 
along the entire inflaton trajectory (which can be inferred from formula \eqref{sgoldappr} and condition \eqref{condApp} in Appendix \ref{regimes}). 
These two observations can be confronted with the results of \cite{Hardwick:2016whe} which consider the contribution to post-inflationary evolution of simple curvaton models for a range of parameters. Our sgoldstino scenario can be argued to map to one where the curvaton  never dominates the energy of the universe, though this question deserves clearly further dedicated study.
  
As already mentioned, we did not include possible effects from the extended SUSY Higgs sector and from the Higgsino, and we further assumed the Bino to be the NLSP. Note that if the neutralino NLSP would be instead mostly Higgsino-like the
bounds from BBN (purple region) and from NLSP decay (the part of the red region on the right of the dashed green line) would be considerably softened since the abundance of the Higgsino-like neutralino is suppressed compared to the
 one of the Bino.
  
Concerning the gravitino abundance, we did not consider possible non thermal production arising during preheating \cite{Giudice:1999am}.
 The results of a recent paper \cite{Hasegawa:2017hgd} seem to indicate that our model, where $h(\Phi)$ is a quadratic function, should not be hindered by such effects, but a dedicated analysis should be performed to reach
 a definite conclusion.
Generically, there can be other effects during the cosmological history of the universe, in particular in the (p)reheating epoch, that would eventually lead to additional gravitino production mechanisms which
 would impact significantly the outcome of our analysis. We leave for future studies a more thorough study of the gravitino problem. 
 However, we argue that our results already hint at the possibility that the gravitino problem can be more easily circumvented in this class of models, due to
 the low reheating temperature, which is an intrinsic property of our model, in particular of the coupling of the inflaton to the supersymmetry breaking mechanism.

\section{Discussion}
\label{conclu}
In this paper we have addressed the issue of the compatibility of nilpotent inflation with low-scale SUSY breaking. The nilpotent approach simplifies many aspects of the supergravity embedding of inflation since the sgoldstino is taken to be integrated out and its dynamics can be neglected.

We consider a class of models in which the sgoldstino is present but has a mass, given by a higher dimensional effective operator, and we have investigated under which conditions this massive sgoldstino is always effectively decoupled from the physics of inflation. Specializing to a field theoretic weakly coupled UV completion of the SUSY breaking sector, which we take to have a non-trivial coupling to the inflaton, we find that the scale of SUSY breaking cannot be as low as one could expect, for instance, in gauge mediated models. This can be intuitively understood as follows: low SUSY breaking scales lead generically to a light sgoldstino, and it becomes more difficult to decouple the latter from inflation physics. The constraints are in practice more complicated, but they can be nicely summarized as in Figure \ref{fig:parameterspace}.

An important remark concerns the implications of our results on the regime of validity of generic nilpotent inflation models. These can encompass various different choices, concerning on one side the inflationary supergravity model, and on the other the type of SUSY breaking dynamics and mediation. Most notably the coupling between two physical set-ups will be important, as in our model. Indeed, in \cite{Dudas:2016eej} a variety of models were considered, both for inflation and for SUSY breaking, but no inflaton-dependence was contemplated in the operator giving the sgoldstino its mass. This resulted in very stringent bounds on the scale of SUSY breaking. Our findings relax these bounds, but nevertheless we cannot explore all of the potential parameter space of the SUSY breaking and mediation scales while staying at weak coupling.

It may well be that generalizing even more the types of models we can push further down the bounds on the gravitino mass and the SUSY breaking scale. Note however that our choice of inflationary sector, the $\alpha$-attractor, is already very flexible in itself. As for the SUSY breaking sector, its effective parametrization as in Section \ref{TheModel} allowed us to derive a crude approximation of the bounds, that we found using the increased precision of the weakly coupled UV complete model for definiteness. One way to try to overcome these bounds is to take strongly coupled SUSY breaking and messenger sectors. Here we could hope to explore the other valid regions of parameter space that open up in the EFT.

The specification of the UV model, and in particular of the couplings between the inflaton and fields involved in SUSY breaking, has a positive side. It has allowed us to discuss in some detail the physics of inflation, including reheating and dark matter abundance bounds, and confront it with collider bounds on superpartner masses. The complementarity of these bounds is manifest in Figure \ref{fig:money}. The result is actually that our model is quite predictive, both for inflation observables (low spectral tilt $n_s$ and reheating temperature) and for collider ones (upper and lower bounds on the gluino and gravitino masses). A different UV completion would certainly change the details, and the outcome of the analysis. We believe however that we have shown how to proceed in such a task. 

To conclude, we would like to convey the message that nilpotent inflation can be compatible with low scales of SUSY breaking only with an increasing number of conditions on its UV completion. We do not seem to reasonably expect it to allow for arbitrarily low scales, i.e.~as for a GMSB scenario with eV-scale gravitino. On the other hand, once a UV completion is specified, such models lead to a complete and complementary characterization of cosmological and collider observables, thus confirming the expectation that inflation and SUSY breaking are intimately tied together.

%%%%%%%%%%%%%%%%%%%%%%%%%%%%%%%%%%%%%%%%%%%%%%%%%%%%%
%  Acknowledgements
%%%%%%%%%%%%%%%%%%%%%%%%%%%%%%%%%%%%%%%%%%%%%%%%%%%%%
\section*{Acknowledgements}

We would like to thank M.~Bertolini, W.~Buchm\"uller, F.~D'Eramo, E.~Dudas, D.~Redigo\-lo, C.~Ringeval, D.~Roest, M.~Tytgat and B.~Vercnocke for discussions.
This research has been supported in part by IISN-Belgium (convention 4.4503.15). R.A.~is
a Senior Research Associate of the Fonds de la Recherche Scientifique-F.N.R.S. (Belgium).
The work of L.H.~is funded by the Belgian Federal Science Policy through the Interuniversity Attraction Pole P7/37.  
A.M.~is supported by the Strategic Research Program High Energy Physics and the Research Council of the Vrije Universiteit Brussel.

%%%%%%%%%%%%%%%%%%%%%%%%%%%%%%%%%%%%%%%%%%%%%%%%%%%%%
%  bibliography
%%%%%%%%%%%%%%%%%%%%%%%%%%%%%%%%%%%%%%%%%%%%%%%%%%%%%
  
\appendix
 
\section{Analysis of sgoldstino VEV and EFT validity}
\label{regimes}
In this appendix we give more details on the analysis of the constraint on the sgoldstino VEV.
As explained in Section \ref{TheModel}, the inflationary trajectory spans the space of $\text{Im}(\Phi)$ with $\text{Re}(\Phi) = \text{Im}(S) = 0$ while the real part of the sgoldstino is given by equation (\ref{vevS}).
The condition $\langle S \rangle \ll \Lambda_{\text{eff}}(\Phi)$ is quite involved to solve analytically given the different terms entering into the expression for the sgoldstino VEV 
 (\ref{vevS}).
 We will discuss various limits for this expression to extract the relevant inequalities such that the validity of the effective theory is guaranteed along the
 entire inflationary trajectory.
 We first analyze the two extrema of the inflaton trejectory.
For small $\varphi$ the sgoldstino VEV is Planck suppressed and scales as $\langle s \rangle \sim \frac{\Lambda_0^2}{M_p}$, hence the EFT condition is trivially satisfied.\footnote{Here we neglect corrections due to a non-zero $m_y$ in the UV theory, which are in any case relevant only at the end of the trajectory.}
For very large inflaton, when $\varphi \sim M_p$, the sgoldstino VEV scales as $\langle s \rangle  \sim \frac{m_h}{m_f} g^2 M_p $. Here we have assumed that $m_h\ll m_f$, since we have indeed the freedom to decouple the scale of the inflaton mass from the scale of inflation itself. We will see instantly that this hierarchy is actually a requirement. 
The condition $\langle s \rangle \ll g \varphi \sim g M_p$ hence gives $g \ll \frac{m_f}{m_h}$ which is trivially satisfied.
The two extrema of the $\varphi$ excursion are hence within the EFT validity range. 

We now have to investigate the rest of the $\varphi$ trajectory.
A good estimate to understand the possible regimes of validity and the constraints on the parameters can be obtained by expanding both numerator and denominator in (\ref{vevS}) in some approximation.
The largest scale of the model is $m_f$ that will set the energy scale of inflation, as illustrated in Section \ref{TheModel}. We then expand the sgoldstino VEV at leading order in 
 $\frac{\rho}{M_p}$, where $\rho$ is any dimensionful parameter except $m_f$, and at second order in $\frac{m_f}{M_p}$,\footnote{This is equivalent to assuming that the expansion parameter is $\epsilon \sim \frac{\rho}{M_p} \sim \frac{m_f^2}{M_p^2}$.} getting
 \be
\label{sgoldappr}
\langle s \rangle =
\frac{  \frac{2 \sqrt{6} f_0^2}{3 M_p}-\frac{2 \sqrt{2} m_f m_h \varphi^2}{3\alpha M_p} }{\frac{2 m_f^2 \varphi^2}{3 \alpha M_p^2} + \frac{ (2 f_0+\frac{m_f \varphi^2}{ M_p})^2}{\Lambda_{0}^2+\frac{g^2}{2} \varphi^2}}\ .
\ee
We should compare this VEV with the scale giving the upper bound on the validity of the effective theory, i.e.~$\sqrt{\Lambda_{0}^2+\frac{g^2}{2} \varphi^2}$.
From now on we omit any numerical $O(1)$ coefficient  (including $\alpha$) for simplicity of the discussion; they can be reinserted easily by inspecting the expression (\ref{sgoldappr}).

There are several critical values along the inflaton trajectory where the sgoldstino VEV changes behavior as a function of $\varphi$. 
These are the values where the different terms in  \eqref{sgoldappr} change from subleading to dominant, which are
\be
\label{regions}
\varphi^2\quad \sim \qquad  \frac{f_0^2}{m_f m_h} \quad , \quad \frac{f_0^2 M_p^2}{\Lambda_0^2 m_f^2} \quad , \quad \frac{f_0 M_p}{m_f} \quad , \quad \frac{\Lambda_0^2}{g^2}\ ,
\ee
where the last one determines also a change in the EFT validity scale.
In determining these critical values we have made the crucial assumption that they are ordered as in \eqref{regions} in increasing size.
Indeed the assumption $\frac{\Lambda_0^2}{g^2} \gg \frac{f_0 M_p}{m_f} $ implies that the next relevant scale for the behaviour in the denominator is $\frac{f_0^2 M_p^2}{\Lambda_0^2 m_f^2}$. This is smaller than 
$\frac{f_0 M_p}{m_f}$ if we further impose $f_0 \ll \frac{m_f \Lambda_0^2}{M_p}$.
The first ordering on the left in \eqref{regions} also requires a further assumption. 
All in all the choice of ordering of the critical values as in \eqref{regions} leads to the following inequalities
\be
f_0 \ll \frac{m_f}{M_p} \Lambda_0^2\ ,  \qquad \qquad  \frac{\Lambda_0^2}{M_p^2} \ll \frac{m_h}{m_f}\ .
\ee
Now we can proceed identifying the inequalities that the parameters should satisfy in order for the sgoldstino VEV to be within the EFT validity regime,
in all the five intervals defined by the critical values in \eqref{regions} from $\varphi \sim 0$ to $\varphi \sim M_p$.

Analyzing all the intervals one finds that the complete set of inequalities
is\footnote{We always assume $g \ll 1$.}
\be
\label{condApp}
f_0  \ll \frac{\Lambda_0^2 m_f}{M_p}\ ,   \qquad   \frac{\Lambda_0^2}{M_p^2} \ll \frac{m_h}{m_f}\ , \qquad \frac{\Lambda_0}{M_p} \gg \frac{m_h}{m_f} \ .
\ee
In particular the last inequality is the one emerging from the analysis of the regimes of validity in the various intervals. Note that it imposes an upper bound on the size of the quadratic term of the inflaton potential, 
resulting in an upper bound on the inflaton mass at the end of inflation, that will have important consequences for the phenomenology.
Moreover, since $\Lambda_0\ll M_p$, it also confirms that $m_h\ll m_f$.

We conclude by observing that by changing the assumptions on the ordering of the various turning points in \eqref{sgoldappr}, one can extend to other regions of parameter space where the EFT is still valid. This has indeed also been checked by randomly scanning over the various parameters. However, we have observed that restricting the scan to EFT parameters compatible with a weakly coupled UV completion as in section \ref{UVmodel}, the parameter space is cut out to the UV version of \eqref{condApp}, that is \eqref{uvcondtraj}. Hence our focus on this region.

\bibliographystyle{JHEP}

\bibliography{myref}  

\end{document}